\documentclass[times,tighten,twocolumn,english]{aastex63}
\usepackage{bm}
\usepackage{lipsum}
\usepackage{physics}
\usepackage{multirow}
\usepackage{rotating}
\usepackage{graphicx}
\usepackage{subfigure}
\usepackage{color}
\usepackage{amsfonts}
\usepackage{mathrsfs}
\usepackage{babel}
\usepackage{amssymb}
\usepackage{wasysym}
\usepackage{verbatim}
\usepackage{multirow}
\usepackage{hyperref}
\usepackage{float}
\usepackage{ulem}

\usepackage{paralist}

\begin{document}

\title{Accretion Disk Size Measurements of Active Galactic Nuclei Monitored by the Zwicky Transient Facility}

\author{Wei-Jian Guo}
\affiliation{Key Laboratory for Particle Astrophysics, Institute of High Energy Physics, Chinese Academy of Sciences, 19B Yuquan Road, Beijing 100049, China; Email:\href{mailto:liyanrong@mail.ihep.ac.cn, wangjm@mail.ihep.ac.cn}{liyanrong@mail.ihep.ac.cn, wangjm@mail.ihep.ac.cn} }
\affiliation{School of Astronomy and Space Science, University of Chinese Academy of Sciences, 19A Yuquan Road, Beijing 100049, China}
\author[0000-0001-5841-9179]{Yan-Rong Li}
\affiliation{Key Laboratory for Particle Astrophysics, Institute of High Energy Physics, Chinese Academy of Sciences, 19B Yuquan Road, Beijing 100049, China; Email:\href{mailto:liyanrong@mail.ihep.ac.cn, wangjm@mail.ihep.ac.cn}{liyanrong@mail.ihep.ac.cn, wangjm@mail.ihep.ac.cn} }
\author{Zhi-Xiang Zhang}
\affiliation{Department of Astronomy, Xiamen University, Xiamen, Fujian 361005, China}
\author[0000-0001-6947-5846]{Luis C. Ho}
\affiliation{Department of Astronomy, School of Physics, Peking University, Beijing 100871, China}
\author[0000-0001-9449-9268]{Jian-Min Wang}
\affiliation{Key Laboratory for Particle Astrophysics, Institute of High Energy Physics, Chinese Academy of Sciences, 19B Yuquan Road, Beijing 100049, China; Email:\href{mailto:liyanrong@mail.ihep.ac.cn, wangjm@mail.ihep.ac.cn}{liyanrong@mail.ihep.ac.cn, wangjm@mail.ihep.ac.cn} }
\affiliation{School of Astronomy and Space Science, University of Chinese Academy of Sciences, 19A Yuquan Road, Beijing 100049, China}
\affiliation{National Astronomical Observatories of China, Chinese Academy of Sciences, 20A Datun Road, Beijing 100020, China}


\begin{abstract}

We compile a sample of 92 active galactic nuclei (AGNs) at $z<0.75$ with $gri$ photometric light curves from the archival data of the Zwicky Transient Facility and measure the accretion disk sizes via continuum reverberation mapping. We employ Monte Carlo simulation tests to assess the influences of data sampling and broad emission lines and select out the sample with adequately high sampling cadences (3 days apart in average) and minimum contaminations of broad emission lines. The inter-band time delays of individual AGNs are calculated using the interpolated cross-correlation function and then these delays are fitted with a generalized accretion disk model, in which inter-band time delays are a power function of wavelength, black hole mass, and luminosity. A Markov-chain Monte Carlo method is adopted to determine the best parameter values. Overall the inter-band time delays can be fitted with the $\tau \ \propto  \lambda^{4/3}$ relation as predicted from a steady-state, optically thick, geometrically thin accretion disk, however, the yielded disk size is systematically larger than expected, although the ratio of the measured to theoretical disk sizes depend on using the emissivity- or responsivity- weighted disk radius. These results are broadly consistent with previous studies, all together raising a puzzle about the ``standard'' accretion disk model.
\end{abstract}


\section{Introduction}

Active galactic nuclei (AGNs) and quasars are those celestial objects that are believed to be powered by mass accretion onto the supermassive black hole (SMBH) residing in each (if not all) galactic center. The formed accretion disks surrounding SMBHs are responsible for the outward angular momentum transportation and huge energy radiation. In sub-Eddington regime, accretion disks are optically thick and geometrically thin and are described by the standard accretion disk model established by the pioneering work of \cite{Shakura1973}. In super-Eddington regime, accretion disks become both optically and geometrically thick and are described by the slim disk model (\citealt{Abramowicz1988,wang1999}). These accretion disk models had made great success in interpreting AGN observational characteristics (see \citealt{Lin1996} for a review), such as the UV/optical continuum emissions (e.g., \citealt{Koratkar1999, Kishimoto2008}) and the broad spectral energy distributions (e.g., \citealt{Sanders1989}). Nevertheless, these models are still not fully understood and a variety of observational tests are clearly warranted (e.g., \citealt{Courvoisier1991, Gierlinski2004, Blaes2007}).

In practice, there are two direct methods to test for accretion disk models in AGNs. One method is through gravitational microlensing (e.g., \citealt{Kochanek2004, Morgan2006}) and the other is through continuum  reverberation mapping (e.g., \citealt{Collier1999, Cackett2007}). In microlensing observations, disk sizes and radial temperature profiles are inferred by analyzing flux variability of multiple lensed images of gravitationally lensed quasars; In continuum reverberation mapping observations, measurements are a bit more straightforward. Inter-band time delays of AGN continuum light curves are interpreted to be the light-traveling time between different emission regions responsible for different wavelength bands. Therefore, such time delays directly provide constraints on physical disk sizes if multiplying the delays with the light speed (\citealt{Collier1999}). The dependence of time delays on wavelength can further determine radial temperature profiles of accretion disks. The both approaches generally found that the measured disk sizes are larger by a factor of about two than the anticipated values from the standard disk model (e.g., \citealt{Morgan2010, Edelson2019, Cornachione2020} and references therein). This conflict had spurred great attention and a large number of studies to measure disk sizes and temperature profiles over the past decades.

Regarding continuum reverberation mapping, there are two major approaches: one approach utilizes space- and/or ground-based telescopes to monitor individual AGNs with intensive cadences and multi-band filters (e.g., \citealt{Fausnaugh2015,Fausnaugh2018,Starkey2016,McHardy2018,Cackett2017, Edelson2019, Cackett2020, Kara2021}). This approach yields high-quality time delays and investigates in details disk structures in individual AGNs, but is apparently limited by telescope resources and sample size. The other approach largely relies on multi-band archival data of time-domain surveys and therefore can explore over large samples (e.g, \citealt{Jiang2017, Mudd2018, Homayouni2019, Yu2020,Lobban2020MNRAS,Homayouni2021, Jha2021}). Despite the relatively poor quality in time delay measurements,  the advantage of large sample sizes still makes this approach essential to draw meaningful constraints on disk structures. In particular, it allows to study the dependence of time delays on AGN properties (such as the luminosity, black hole mass, and accretion rate), therefore providing additional tests for accretion disk models.

In this work, we compile a sample of 92 AGNs from the Zwicky Transient Facility (ZTF) archives (\citealt{Graham2019}) with sufficient data quality of $g$-, $r$- and $i$-band photometric light curves to determine the inter-band time delays. We then use those time delays to investigate their dependence on various physical disk parameters, including black hole mass, mass accretion rate, and radial temperature profiles, aiming at measuring accretion disk structures. Compared to previous time-domain survey data, the ZTF data have a relatively more homogeneous and higher cadence (about 3 days apart) and lower measurement noises (typically $\lesssim3\%$). These properties enable us to place more tight constraints on disk models. 

The paper is organized as follows. Section \ref{sec2} describes the data and sample selection based on the 3rd ZTF public data release. Section \ref{sec_disk_all} generalizes the standard accretion disk model, in which inter-band time delays are formulated as a power function of wavelength, black hole mass, and luminosity. In Section~\ref{sec_disk_meausre}, we use the obtained time delays to measure disk structures. Discussions  and conclusions are given in Sections \ref{sec_discussion} and \ref{sec_conclusion}, respectively. Throughout the paper, we use a $\Lambda$CDM cosmology with $H_{0}= \rm{67\ km\  s^{-1}\  Mpc^{-1}}$, $\Omega_{\Lambda}= 0.68$, and $\Omega_{m}= 0.32$ (\citealt{Planck2020}).

\begin{deluxetable*}{ccccccccccc}
\tablecolumns{10}
\tabletypesize{\footnotesize}
\tabcaption{\centering Sample Properties \label{tab1}}
\colnumbers
\tablehead{
\colhead{Object Name} &
\colhead{R.A.} &
\colhead{Dec.} &
\colhead{$z$} &
\colhead{$\log(M_{\bullet}/M_{\odot})$} &
\colhead{$\log(L_{5100}/\rm erg\ s^{-1})$} &
\colhead{$\dot{\mathscr{M}}$} &
\colhead{$r_{gr}$} &
\colhead{$r_{gi}$} &
\colhead{$\tau_{gr}$} &
\colhead{$\tau_{gi}$}\\
& (deg) & (deg) & & & & & & & (day) & (day)
}
\startdata
SDSS J002029.37+192153.0 & 5.1224 & 19.3647 & 0.627 & 8.56$\pm${0.17} & 45.12$\pm${0.05} & 1.09 & 0.82 & 0.83 & 2.54$^{+4.42}_{-4.41}$ & 2.55$^{+7.95}_{-5.93}$\\
SDSS J135955.65+504539.5 & 209.9819 & 50.7610 & 0.365 & 7.08$\pm${0.17} & 44.67$\pm${0.03} & 214.59 & 0.90 & 0.79 & -0.58$^{+3.50}_{-3.30}$ & 4.06$^{+6.30}_{-8.91}$\\
SDSS J140201.84+443558.1 & 210.5077 & 44.5995 & 0.406 & 7.80$\pm${0.15} & 44.70$\pm${0.04} & 8.66 & 0.83 & 0.65 & -1.38$^{+4.13}_{-4.28}$ & 6.56$^{+7.94}_{-11.91}$\\
SDSS J140604.24+572956.6 & 211.5177 & 57.4990 & 0.326 & 8.54$\pm${0.14} & 44.25$\pm${0.05} & 0.06 & 0.87 & 0.80 & -1.68$^{+2.56}_{-2.82}$ & -2.03$^{+8.88}_{-6.15}$\\
SDSS J141740.44+381821.1 & 214.4185 & 38.3059 & 0.450 & 7.76$\pm${0.15} & 44.50$\pm${0.04} & 5.40 & 0.89 & 0.68 & -4.10$^{+4.37}_{-3.08}$ & -2.30$^{+10.46}_{-11.24}$\\
SDSS J141956.64+583501.3 & 214.9860 & 58.5838 & 0.271 & 7.92$\pm${0.16} & 44.32$\pm${0.04} & 1.35 & 0.85 & 0.69 & 4.85$^{+2.44}_{-2.57}$ & 6.86$^{+4.89}_{-6.03}$\\
\enddata
\tablecomments{Columns: (1) name, (2) right ascension,  (3) declination, (4) redshift, (5) black hole mass, (6) optical luminosity at 5100~\AA, \  (7) dimensionless accretion rate, (8) maximum cross-correlation coefficient between $g$- and $r$-band, (9) maximum cross-correlation coefficient between $g$- and $i$-band, (10) time delay between $g$- and $r$-band in the observed-frame, and (11) time delay between $g$- and $i$-band in the observed-frame.\\
(This table is available in its entirety in machine-readable form.)}
\end{deluxetable*}

\section{The Data and Sample} \label{sec2}

\subsection{Photometric Light Curves and Spectra}
The ZTF is a fully-automated time-domain survey aimed at a systematic exploration of the optical transient sky.  It uses a 48-inch Schmidt telescope located at the Palomar Observatory, which is equipped with a wide-field camera  (47 square degrees field of view) to photometrically scan the entire northern visible sky at a rate of $\sim$3760 square degrees per hour (\citealt{Bellm2019}). There are three custom filters ($g$-, $r$-, and $i$-band) that cover a wavelength range approximately from  4100~\AA\ to 8700~\AA\  \citep{Masci2019}. The typical sampling interval is about three days (\citealt{Graham2019}). 

\begin{figure*}[tp!]
\centering 
\includegraphics[width=0.8\textwidth]{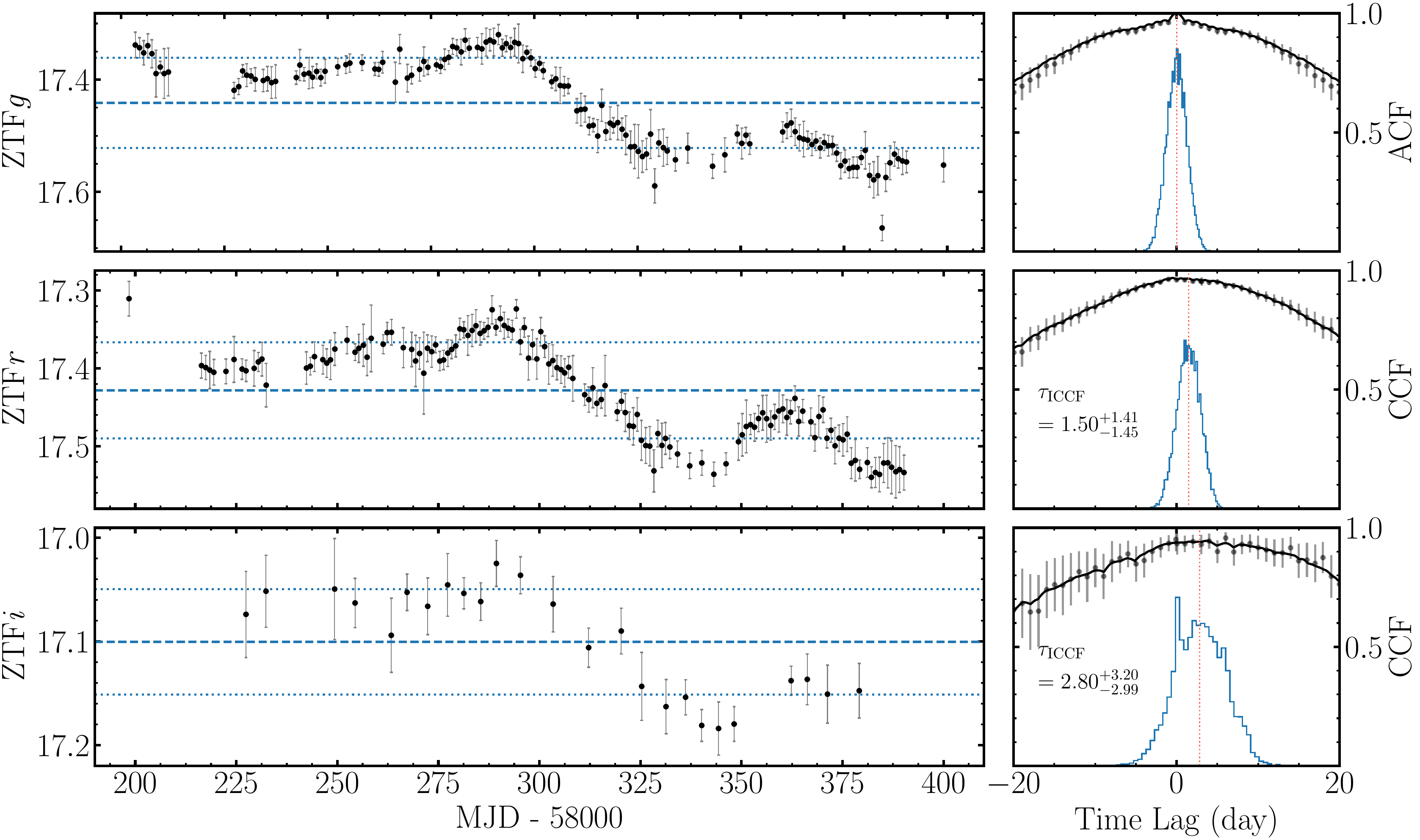}
\caption{
(Left) $g$-, $r$- and $i$-band light curves of an AGN selected from the ZTF DR3. Horizontal dashed and dotted lines represent the mean and $1\sigma$ fluxes of the light curves, respectively. (Right) from top to bottom panels are the ACF of the $g$-band light curve and the ICCFs of $r$- and $i$-band light curves with respect to $g$-band one, respectively. In each panel, the histogram represents the cross-correlation centroid distribution.  Grey points with error bars show the ZDCFs.
}
\label{ccf}
\end{figure*}

We used the third ZTF public data release (DR3\footnote{\url{https://www.ztf.caltech.edu}.}). When querying the data archives, we switched on the flag $\mathtt{catflags} = 0$ to filter out ``bad-quality'' data points (\citealt{Masci2019}). There are usually multiple exposures per night. We combined those exposures  in average on the same night. The temporal baseline of DR3 spans two years (MJD 58200-58900) with a seasonal gap of about three months. The data in the first year have a relatively higher sampling rate, therefore, we only use the first-year data (MJD 58200-58400) for subsequent analysis. For the sake of illustration, Figure~\ref{ccf} plots the $g$-, $r$-, $i$-band light curves of a selected AGN.

To determine the black hole mass, optical luminosity at 5100~\AA\,, and broad emission line properties, we cross-matched the preselected sample with the quasar catalog of the 14th data release of the Sloan Digital Sky Survey\footnote{\url{https://www.sdss.org/}} (DR14Q; \citealt{Paris2018}). The redshift is limited to $z<0.75$ to ensure the presence of broad H$\beta$ lines in spectra, with which the virial black hole mass can be estimated. There are about ten thousand AGNs at a redshift $z<0.75$ with three-band data. Below we further censor those AGNs to select out ones with data quality sufficient to measure reliable inter-band delays. 

\subsection{AGN Properties}\label{sec2.2} 
\begin{figure*}[th!]
\centering 
\includegraphics[scale=0.5]{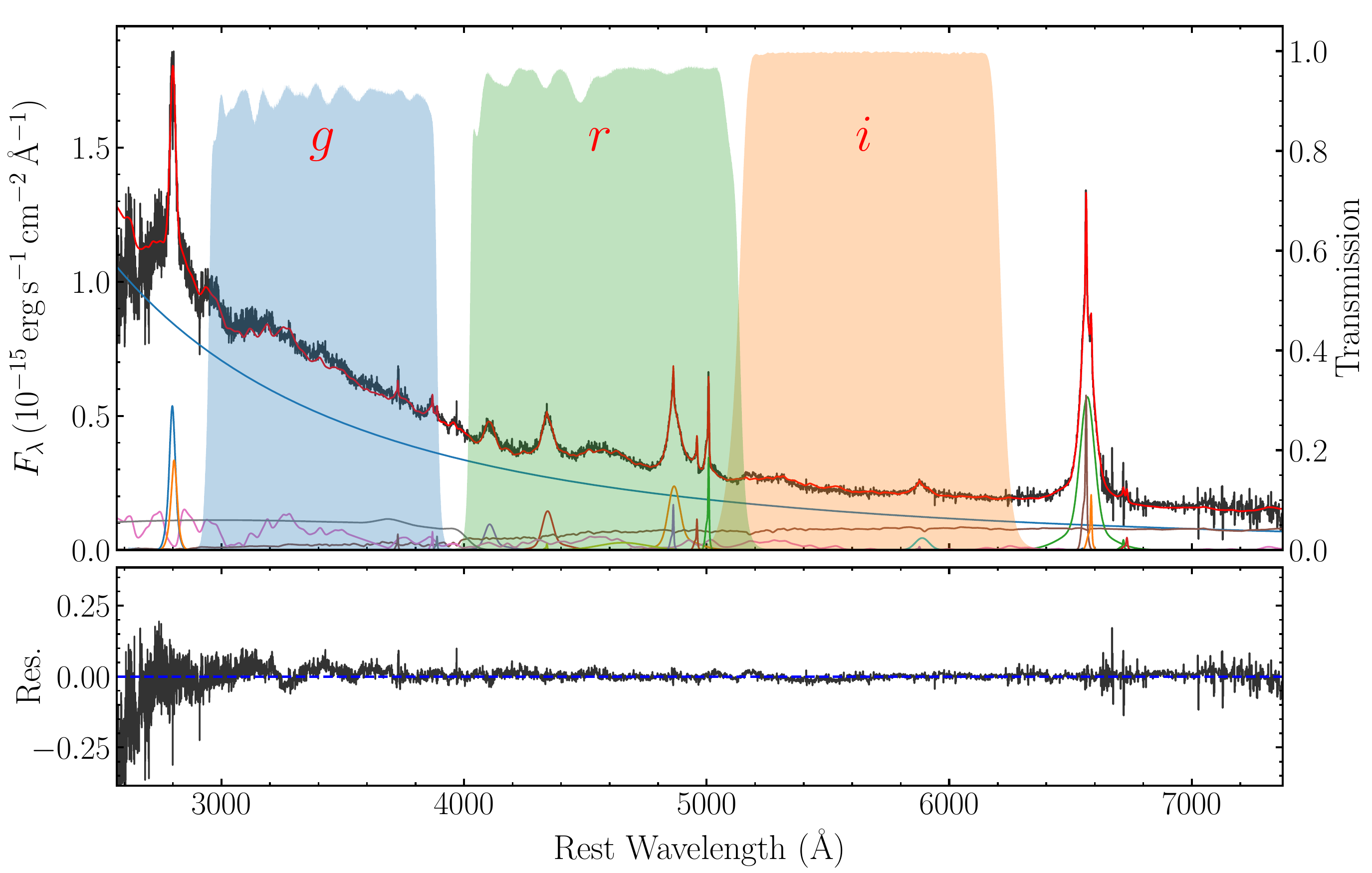}
\caption{An example for our spectral decomposition. The top panel shows the fitting results and the bottom panel shows the residuals. Blue, green, yellow  shaded areas represent the transmissions of $g$-, $r$-, and $i$-band filters, respectively. }
\label{spectrum}
\end{figure*}

We employ a spectral decomposition scheme following \cite{Barth2013} and \cite{Hu2015} to derive the luminosity at 5100~\AA\, and emission line properties. We adopt the \texttt{DASpec} software\footnote{\url{https://github.com/PuDu-Astro/DASpec}}, which uses the Levenberg-Marquardt method for the chi-square optimization. The spectral components include a featureless power-law for the continuum, an Fe II template from \cite{Boroson1992}, double Gaussians for the prominent broad emission lines (such as H$\alpha$, H$\beta$, H$\gamma$, \ion{He}{2}, etc), a single Gaussian for the narrow emission lines (such as [\ion{O}{3}]$\lambda\lambda$4959, 5007, [\ion{O}{3}]$\lambda$4363, etc), a Balmer continuum, 
 and a host galaxy template.  For simplicity, the host galaxy template is chosen to be a single stellar population model with an instantaneous burst of 11 Gyr and a metallicity of $Z= 0.05$ from \cite{Bruzual2003}.
We derive the 5100~{\AA} luminosity from the power-law component, which means that the host galaxy component is naturally 
excluded.
With the above spectral decomposition, we  can also estimate the flux contributions of emission lines in each filter's bandpass. Figure~\ref{spectrum} shows an example of our spectral decomposition and the transmissions of the three ZTF filters. 

Given with the 5100~{\AA} luminosity, the black hole mass is  estimated by combining the size-luminosity ($R_{\rm{H}\beta}$-$L_{5100}$) scaling relation and H$\beta$ line widths ($V_{\rm H\beta}$) from the single-epoch spectra, namely,
\begin{equation}
M_{\bullet}=\it f_{\rm{{BLR}}}\frac{V^{\rm{2}}_{\rm{H\beta}}R_{\rm{H\beta}}}{G},
\label{eq_mass}
\end{equation}
where $G$ is the gravitational constant, $f_{\rm BLR}$ is the virial factor, and $R_{\rm{H\beta}}$ is determined from the $R_{\rm{H}\beta}$-$L_{5100}$ relation. We use the $R_{\rm{H}\beta}$-$L_{5100}$ relation compiled by \cite{Du2019}, which, compared to that compiled by \cite{Bentz2009,Bentz2013}, covers both sub- and super-Eddington AGNs and also takes into account the accretion rate,
\begin{equation}
\log(R_{\rm H\beta}/{\rm ld}) = \alpha_0 + \beta_0 \log{\ell}_{\mathrm{44}}+\gamma_0 \it \mathcal{R}_{\rm{Fe}},
\end{equation}
where $\ell_{44}=L_{5100}/10^{44} \rm{erg\ s}^{-1}$ and $\mathcal{R}_{\rm{Fe}}$ is the flux ratio between \ion{Fe}{2} and H$\beta$ lines, which is strongly correlated with the accretion rate (e.g., \citealt{Du2016}).
Here, the coefficient values are $\alpha_{0}=1.65\pm0.06$,  $\beta_{0}=0.45\pm0.03$, $\gamma_{0}=-0.35\pm0.08$ (see \citealt{Du2019} for a detail). We use the full width at half maximum (FWHM) as the measure of H$\beta$ line width and adopt a virial factor  ${f_{\rm BLR}}=1.12\pm0.31$ \citep{Woo2015}. 

In the following analysis, we also define the dimensionless accretion rate as \citep{Du2014}
\begin{equation}
\mathscr{\dot M} = 20.1\left( \frac{\ell_{44}}{\cos i}\right)^{3/2}m_{7}^{-2},
\end{equation}
where $m_{7}=M_{\bullet}/10^{7}M_\odot$ and $i$ is the inclination angle of the accretion disk. We assume an averaged $\cos i\approx0.75$, which corresponds to a typical inclination of AGNs with broad emission lines  (e.g., \citealt{Fischer2014,Du2016}).

\subsection{Time Delay Measurements} 
\label{sec_time_delay}
We first use the interpolated cross-correlation method to measure inter-band time delays between $r/i$-band  and $g$-band light curves (\citealt{Gaskell1986,Gaskell1987,Peterson1993}). The time delay is assigned the centroid of the interpolated cross-correlation functions (ICCFs) above 80$\%$ of the peak value. The associated uncertainties are estimated by 15.87$\%$ and 84.13$\%$ quantiles of the cross-correlation centroid distribution (CCCD), which is generated by the `` Flux Randomization/Random Subset  Selection (FR/RSS)'' method \citep{Peterson1998,Peterson2004}.
The left panels of Figure~\ref{ccf} show the autocorrelation function (ACF), ICCF, and CCCD of the light curves for a selected AGN. The maximum cross-correlation coefficient $r_{\rm max}$ and the measured  time  delays $\tau_{gr}$ and $\tau_{gi}$ are also listed in Table~\ref{tab1}.  For the sake of illustration, we also superimpose the $z$-transformed discrete correlation functions (ZDCFs; \citealt{Alexander1997, Alexander2013arXiv}) in Figure~\ref{ccf}. The ZDCF method does not require  interpolation to the light curves but is sensitive to 
binning in the correlation space and does not perform well for low-cadence data. We therefore do not use the ZDCF method for subsequent 
analysis.

\begin{figure*}[th!]
\centering
\includegraphics[width=0.95\textwidth]{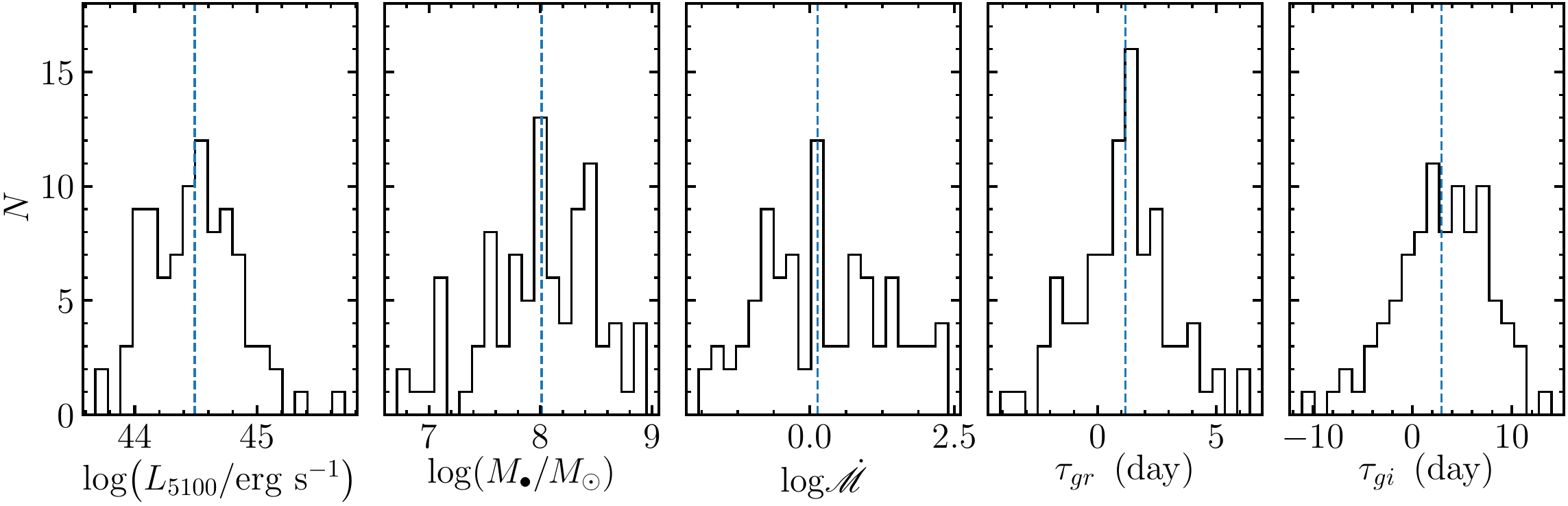}
\caption{The histograms for luminosity at 5100\AA, black hole mass, dimensionless accretion rate, inter-band time delays $\tau_{gr}$ and $\tau_{gi}$ (in the observed frame)  from left to right panels. The blue lines represent the median values.}
\vspace{+0.5cm}
\label{fig_distribution}
\end{figure*}

In Appendix~\ref{sec_mica},  we use a model-independent approach, von Neumann estimator, to cross-check the ICCF results. The Von Neuman estimator is based on the regularity of randomness of data points in light curves (see \citealt{chelouche2017} for detail) and does not 
require an interpolation or binning manipulation. In addition,
we also employ the  Bayesian analysis package \texttt{MICA}\footnote{\texttt{MICA} is available at \url{https://github.com/LiyrAstroph/MICA2}.} (\citealt{Li2016}). \texttt{MICA} uses the damped random walk model to describe the AGN variability (\citealt{Kelly2009, Zu2013,Kasliwal2015}) and a family of displaced Gaussians to approximate the transfer function between the driving and echoed light curves. \texttt{MICA} obtains the posterior distributions of the model parameters using the Markov-chain Monte Carlo technique with a diffusive nested sampling algorithm (\citealt{Brewer2011}; implemented by the package \texttt{CDNest}\footnote{\texttt{CDNest} is available at \url{https://github.com/LiyrAstroph/CDNest}.}). For simplicity, we only use one Gaussian in \texttt{MICA} and assign the time delay as the Gaussian center. 
 As shown in Appendix~\ref{sec_mica}, we find general consistency among the ICCF, von Neumann estimator, and MICA methods.

\subsection{Influences of Broad Emission Lines and Light Curve Sampling}\label{sec_test}
There are two factors that may influence the obtained inter-band time delays. First, the presence of broad emission lines will cause shorter or longer time delays than the realistic values, depending on the flux contributions of the broad emission lines in each filter band-pass  (see Section \ref{sec_dis_bel} below). Second, since the sampling intervals ($\sim$ 3 days) are comparable with or even longer than the anticipated time delays of accretion disks, the sampling rates and gaps in the light curves might be important and also affect the observed time delays. Below we employ Monte Carlo simulations to test these influences and select AGNs with minimized contaminations. 

\begin{figure}
    \centering 
    \includegraphics[width=0.48\textwidth]{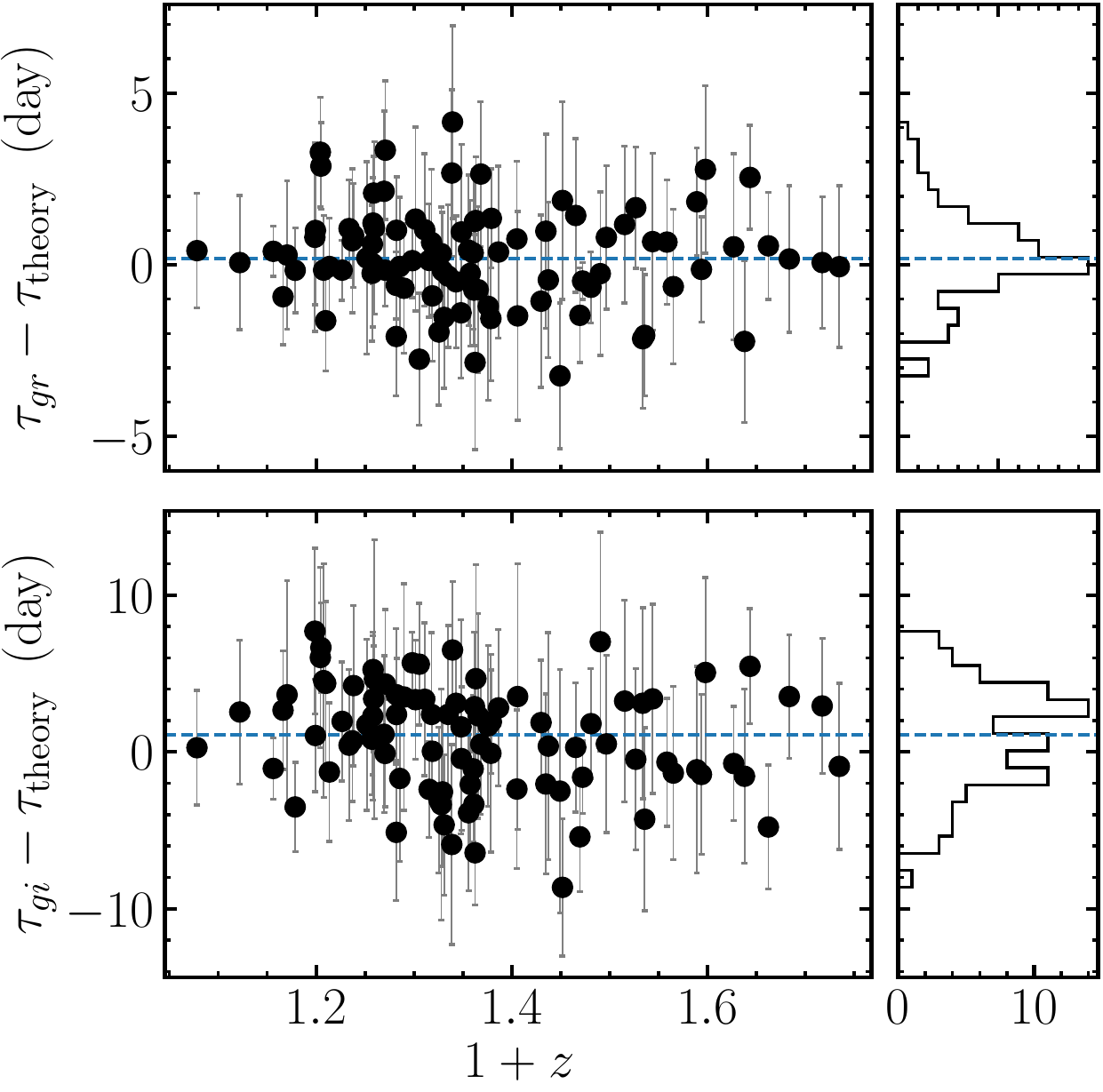}
    \caption{
    The differences of the rest-frame time delays of (top) $r$-band and (bottom) $i$-band with respect to $g$-band light curves compared to the corresponding theoretical delays based on the standard accretion disk model with $\chi=5.04$.
    }
    \label{lag_dis}
\end{figure}

In simulations, we treat the $g$-band light curve as a realistic continuum light curve without contaminations of broad emission lines for the sake of simplicity. We first employ the damped random walk model (\citealt{Kelly2009,Zu2013}) to reconstruct the continuum light curves. Using the black hole mass and luminosity obtained from the spectral decomposition, we calculate the theoretical time delays of $r$- and $i$- bands with respect to $g$-band based on the standard accretion disk model (see Section~\ref{sec_disk_model} below). We generate mock $r/i$-band continuum light curves by shifting the reconstructed $g$-band light curve with the theoretical time delays and then interpolate them onto the observed time epochs. To generate mock light curves of broad emission lines, we assume that all emission lines have the same time delays as derived from the $R_{\rm{H}\beta}$-$L_{5100}$ scaling relation and the corresponding transfer function is Gaussian with a width set to a quarter of the time delay. We convolve the reconstructed $g$-band continuum light curve with this Gaussian transfer function and scale the obtained light curve to force the same flux contribution in $g/r/i$ band as in the spectral decomposition. We finally add up light curves of the continuum and emission lines to create the contaminated mock $g$-, $r$-, and $i$-band light curves and assign the uncertainties by the relative errors of the observed light curves. We calculate the time delays of mock light curves using the same interpolated cross-correlation method as applied to the observed light curves. In Appendix~\ref{app_sim}, we show examples of mock light curves and the corresponding ICCF results.

By repeating the above procedure 5000 times, we obtain time delay distributions of $r$ and $i$ bands for each AGN in the sample. We calculate means from the distributions and assign their uncertainties by the 68.3\% confidence intervals. We first discard those AGNs with mock uncertainties larger than 12 days, which indicates poor data quality. From the remaining AGNs, if the mean delay is consistent  with the input value within the mock uncertainties, we mark that the corresponding observed time delay is reliable and the influences of emission lines and/or data sampling are minimized.

\subsection{Sample Selection Criteria}
\label{sec_criteria}
We design the following criteria to select those AGNs that have sufficiently good data quality to reliably measure inter-band time delays.

\begin{itemize}
\item Both $g$- and $r$-band light curves have epochs $N_{g,r}\geq 60$ over the time baseline of $\sim$200 days. Because the $r$-band sampling rate is relatively low, we relax the criteria to epochs $N_{i}\geq 20$.
    
\item The $g$-, $r$- and $i$-band light curves have statistically significant variability with $\sigma^2 \geq \Delta^2$,
where
\begin{eqnarray}
\sigma^2=\sum_{j=1}^n{\frac{({F}_{j}-\langle F\rangle)^2}{N-1}},
\ \Delta^2=\sum_{j=1}^n\frac{\Delta^2_{j}}{N}.
\end{eqnarray}
where $N$ is the epoch number, $F_j$ and $\Delta_{j}$ are the flux and measurement error at the $j$-th epoch, respectively, and $\langle F\rangle$ is the averaged flux.

\item The ICCFs of the $r$- and $i$-band light curves are calculated with respect to the $g$-band light curve. The maximum cross-correlation coefficients are restricted to $r_{\rm max} > 0.8 $ for $r$-band and $r_{\rm max} > 0.6 $ for $i$-band.
     
\item The ACF FWHM of the $g$-band light curve is restricted to $< 80$~days. This is devised to discard those light curves with little variation patterns so that the cross-correlation analysis usually has  difficulty in determining the time delay.
     
\item As described in \ref{sec_test}, the uncertainties of time delays for mock light curves are restricted to $< 12$ days and the differences between the mock time delay and the input values are within the uncertainties.
\end{itemize}

We finally obtain a sample of 92 AGNs. In Table~\ref{tab1}, we summarize the basic properties of the sample. In Figure~\ref{fig_distribution}, we plot the distributions of 5100~\AA\ luminosity ($L_{\rm 5100}$), black hole mass ($M_\bullet$), dimensionless accretion rate ($\mathscr{\dot M}$),  and inter-band time delays ($\tau_{gr}$ and $\tau_{gi}$) of the sample. Overall, the black hole mass ranges from $10^7$ to $10^9M_\odot$, the dimensionless accretion rate ranges from $\sim$0.01 to 100, and the 5100~\AA\ luminosity ranges from $10^{44}$ to $10^{45}~\rm erg~s^{-1}$. The inter-band time delays are on the order of days and $\tau_{gr}$ is generally smaller than $\tau_{gi}$  as expected. In Figure~\ref{lag_dis}, we show the residuals between the observed inter-band time delays and theoretical  delays  based  on  the standard accretion disk model with $\chi= 5.04$ (see Section~\ref{sec_disk_all} below).

\begin{figure*}[th!]
\centering
\includegraphics[width=0.9\textwidth]{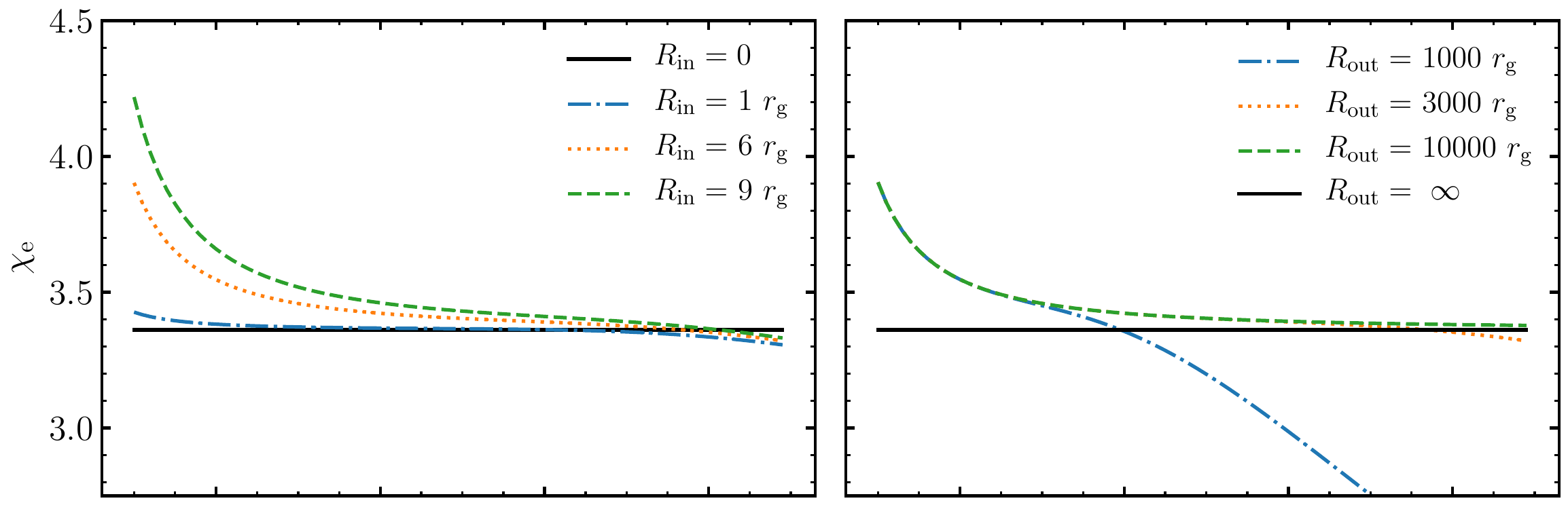}\vspace*{0.01cm}
\includegraphics[width=0.9\textwidth]{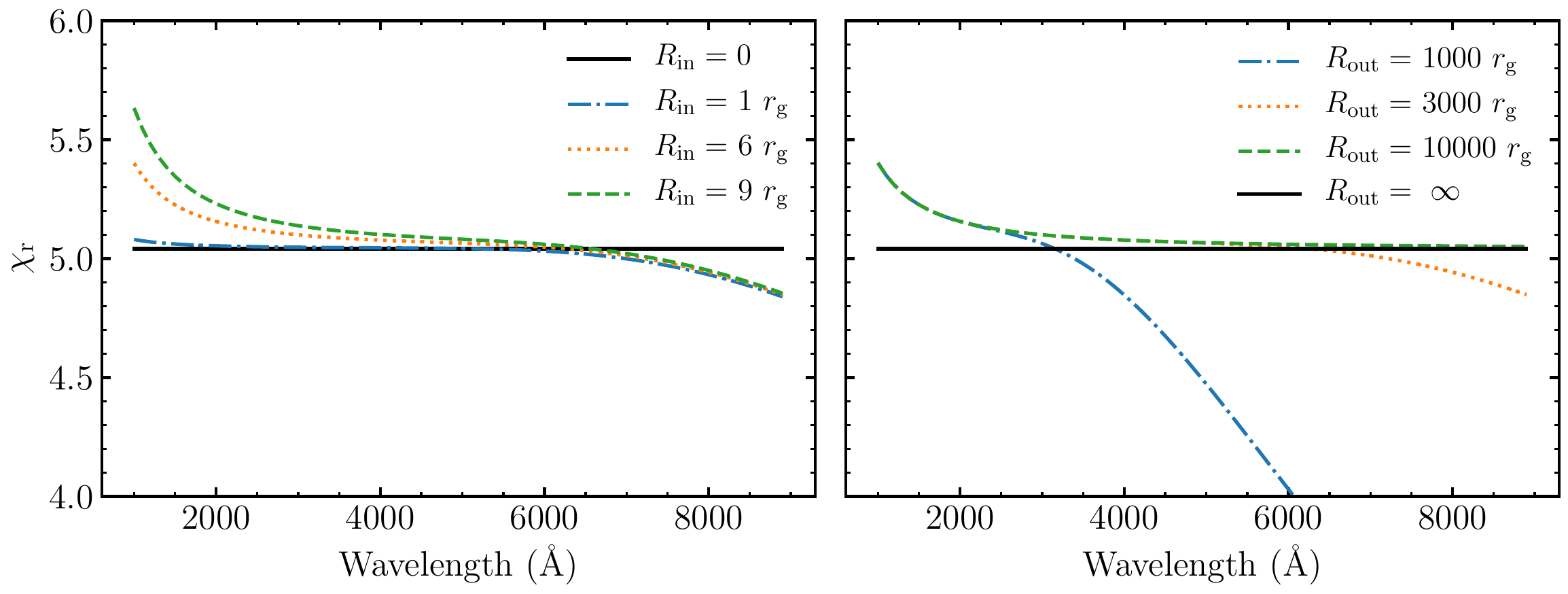}
\caption{
The emissivity-weighted factor $\chi_{\rm e}$ (top) and responsivity-weighted factor $\chi_{\rm r}$ (bottom) with wavelength for different inner and outer disk radius  under the standard disk model. In left panels $R_{\rm{out}}=3000 r_{\rm{g}}$ and in right panels $R_{\rm{in}}=6 r_{\rm{g}}$. The black line in each panel corresponds to $R_{\rm in}\rightarrow0$ and $R_{\rm out}\rightarrow\infty$.
}
\label{fig_factor}
\end{figure*}

\section{Accretion Disk Models}
\label{sec_disk_all}

\subsection{The Standard Accretion Disk Model}
\label{sec_disk_model}
In the standard accretion disk model, by assuming a Keperian rotation, the effective temperature $T$ at radius $R$ can be obtained by equating the radiated flux per disk's surface area to the viscous dissipation rate
\begin{equation}
\label{eq_ssdisk}
\sigma T^{4}=\frac{3GM_{\bullet}\dot{M}}{8\pi R^{3}},
\end{equation}
where $\sigma$ is the Stefan-Boltzmann constant and $\dot{M}$ is the mass accretion rate. To establish the relation between the wavelength and time delay, we perform the following manipulations. At a given wavelength $\lambda$, the temperature is 
\begin{equation}
T=\frac{hc}{k\lambda},
\end{equation}
which corresponds to a radius using Equation~(\ref{eq_ssdisk})
\begin{eqnarray}
R_{\lambda} &=& \left(\frac{3G k^4}{8\pi \sigma h^4 c^4 }\right)^{1/3} M_\bullet^{1/3}\dot{M}^{1/3}\lambda^{4/3}\nonumber\\
& = & \left(\frac{3G C_{\rm bol}k^4}{8\pi \sigma \eta h^4 c^6 }\right)^{1/3} M_\bullet^{1/3}L_{\rm 5100}^{1/3}\lambda^{4/3},
\label{eq_rl}
\end{eqnarray}
where $h$ is the Planck constant, $k$ is the Stefan-Boltzmann constant, $C_{\rm bol}=9.26$ is the bolometric correction factor for the 5100~\AA\, luminosity $L_{5100}$ (\citealt{shen2008, sanchez2018}), 
$\eta=0.1$ is the radiative efficiency, and $\dot M$ is related to $L_{5100}$ as 
\begin{equation}
\dot M = \frac{C_{\rm bol} L_{5100}}{\eta c^2}.
\end{equation}

The observed time delay represents the weighted radius of regions that radiate photons at the wavelength $\lambda$, namely, 
\begin{equation}
c\tau_{\rm \lambda} = \langle R \rangle_{\rm \lambda}=\chi R_{\rm \lambda}.
\label{eq_tau}
\end{equation}
where $\langle R \rangle_{\rm \lambda}$ represents the weighted radius and $\chi$ is a multiplicative factor. Combining Equations~(\ref{eq_rl}) and (\ref{eq_tau}) gives
\begin{equation}
c\tau_{\rm \lambda} = \left(\frac{3G C_{\rm bol} k^4 X^4}{8\pi \sigma \eta h^4 c^6 }\right)^{1/3} M_\bullet^{1/3}L_{5100}^{1/3}\lambda^{4/3},
\label{eq_tau2}
\end{equation}
where we use the conventional quantity in the literature
\begin{equation}
X \equiv X_{\rm ss} = \chi^{3/4}.
\end{equation}
Here,  the subscript ``ss'' represents the standard accretion disk and $X$ (or $\chi$) is introduced to account for the fact that a broad range of disk radius emits photons of a given wavelength. Below we will show how to calculate the values of $\chi$ and $X$  (see Equations~\ref{eq_chie0} and \ref{eq_chir0}). 

In reality, we measure time delays with respect to a reference wavelength, say, e.g., $\lambda_0$. According to Equation~(\ref{eq_tau2}), the observed time delay is then written as 
\begin{equation}
\tau(\lambda)= \tau_{\rm d}(\lambda_0) \left[\left(\frac{\lambda}{\lambda_0}\right)^{4/3}-1 \right],
\end{equation}
where $\tau_{\rm d}$ is the time delay at the wavelength $\lambda_0$, depending on black hole mass and luminosity. For a large sample of AGNs with different redshifts, black hole masses, and luminosities, it is more convenient to set the reference wavelengths to a fixed common rest-frame wavelength ($\lambda_0$) and normalize the coefficient $\tau_{\rm d}$ with respect to a fixed black hole mass and luminosity. As such, the observed time delay (in the observed frame) between two wavelength bands (say, $\lambda_1$ and $\lambda_2$; also in the observed frame) for an AGN at a redshift $z$ is then given by
\begin{equation}
\tau_{\rm obs} = \tau_{\rm d} M_8^{1/3} \ell_{\rm 44.5}^{1/3} (1+z)^{-1/3} \left[\left(\frac{\lambda_{1}}{\lambda_{0}}\right)^{4/3}-\left(\frac{\lambda_{2}}{\lambda_{0}}\right)^{4/3}\right],
\label{eq_tau3}
\end{equation}
where $M_8=M_\bullet/10^8M_\odot$, $\ell_{44.5}=L_{5100}/10^{44.5}~\rm erg~s^{-1}$, and $\tau_{\rm d}$ denotes the time delay at the wavelength $\lambda_0$ (in the rest frame) with $M_\bullet=10^8M_\odot$ and $L_{5100}=10^{44.5}~\rm erg~s^{-1}$.
 For the standard accretion disk, $\tau_{\rm d}$ is given by
\begin{eqnarray}
\tau_{\rm d}& =& \frac{1}{c} \left(\frac{3G C_{\rm bol} k^4 X_{\rm ss}^4}{8\pi \sigma \eta h^4 c^6 }\right)^{1/3}\nonumber\\
& & \times (10^8M_\odot)^{1/3}(10^{44.5}~{\rm erg~s^{-1}})^{1/3}\lambda_0^{4/3}\nonumber\\
& &= 0.67X_{\rm ss}^{4/3}(\lambda_0/7000\rm \AA)^{4/3}~light\mbox{-}day.
\label{eq_size1}
\end{eqnarray}

\subsection{Generalizing the Standard Accretion Disk Model}
Equation~(\ref{eq_tau3}) illustrates the time delay as a function of black hole mass, luminosity, and wavelength under the standard accretion disk model.  We generalize Equation~(\ref{eq_tau3}) into a form of (see also \citealt{Homayouni2019})
\begin{equation}
\tau_{\rm obs} = \tau_{0} (1+z)^{1-\beta} M_8^{\gamma}\ell_{44.5}^{\delta}\left[\left(\frac{\lambda_{1}}{\lambda_{0}}\right)^{\beta}-\left(\frac{\lambda_{2}}{\lambda_{0}}\right)^{\beta}\right],
\label{eq_tau4}
\end{equation}
where $\tau_0$, $\beta$, $\gamma$, and $\delta$ are free parameters. The physical meanings of the above parameters can be understood as follows. Equation~(\ref{eq_tau4}) corresponds to a relation between the temperature and radius as
\begin{equation}
\sigma T^{4}=\left(\frac{\tau_0}{\tau_{_{\rm X}}}\right)^{4/\beta}\frac{3GM_\bullet^{4\gamma/\beta}\dot{M}^{4\delta/\beta}}{8\pi R^{4/\beta}},
\label{eq_gendisk}
\end{equation}
where $\tau_{_{\rm X}}$ is given by
\begin{eqnarray}
\tau_{_{\rm X}} & = & \frac{1}{c}\left(\frac{3Gk^4 X^4}{8\pi \sigma  h^4 c^4 }\right)^{\beta/4}
\left(\frac{C_{\rm bol}}{\eta c^2}\right)^\delta\nonumber\\
& & \times (10^8M_\odot)^{\gamma}(10^{44.5}~{\rm erg~s^{-1}})^{\delta}\lambda_0^{\beta},
\label{eq_size2}
\end{eqnarray}
and 
\begin{equation}
X = \chi^{1/\beta}.
\end{equation}
As can be seen, $\tau_0$ measures the disk size, $\beta$ controls the temperature profile with radius, and $\gamma$ and $\delta$ control the dependence of the radiated flux per disk's surface area on black hole mass and accretion rate, respectively. By setting $\tau_0=\tau_{\rm d}$, $\beta=4/3$, $\gamma=1/3$, and $\delta=1/3$, we return to the case of the standard accretion disk model  and $\tau_{\rm x} = \tau_{\rm d}$. It is worth stressing that for a general set of $\beta$, $\gamma$, and $\delta$, $\tau_{\rm x}$ no longer has the dimension of time and therefore does not indicate the disk size. Below we use this generalized Equation~(\ref{eq_tau4}) to fit the observation data. 
\begin{deluxetable}{cccccc}
\renewcommand\arraystretch{1.2}
\tabletypesize{\footnotesize}
\tablewidth{0.48\textwidth}
\tablecolumns{6}
\tabcaption{The multiplicative factors $\chi$ and $X$ in the case of $R_{\rm in}=0$ and $R_{\rm out}=\infty$. \label{tab_xi}}
\tablehead{
\colhead{} &
\multicolumn{2}{c}{~~~~Emissivity Weighting~~~~} & &
\multicolumn{2}{c}{~~~~Responsivity Weighting~~~~} \\\cline{2-3} \cline{5-6}
\colhead{~~~~~~$\beta$~~~~~~} & \colhead{~~$\chi$~~} & \colhead{~~$X$~~} & & \colhead{~~$\chi$~~} & \colhead{~~$X$~~}
}
\startdata 
1    &  1.46  & 1.46 & & 2.19 & 2.19  \\
4/3  &  3.36  & 2.49 & & 5.04 & 3.37  \\
5/3  &  7.81  & 3.43 & & 11.71 & 4.37  \\
2    &  18.80 & 4.36 & & 28.19  & 5.38
\enddata
\end{deluxetable}

\subsection{The Multiplicative  Factor $\chi$}
\label{sec4.2}
In the literature, there are two approaches for calculating the weighted radius $\langle R \rangle_{\rm \lambda}$. One approach is emissivity weighting and the other is responsivity weighting (\citealt{Starkey2016, Tie2018}). For emissivity weighting, the multiplicative factor is directly given by
\begin{equation}
\chi \equiv \chi_{\rm e} =
\frac{1}{R_{\lambda}}\frac{\int_{R_{\rm in}}^{R_{\rm out}}B_\lambda(T)R^2 d R}{\int_{R_{\rm in}}^{R_{\rm out}}B_\lambda(T)R d R},
\label{eq_chie0}
\end{equation}
where the subscript ``e'' represents emissivity weighting, $R_{\rm in}$ and $R_{\rm out}$ are the inner and outer radius of the accretion disk, respectively, and $B_\lambda(T)$ is the Planck function of a temperature $T$ at a wavelength $\lambda$,
\begin{equation}
B_\lambda(T) = \frac{2hc}{\lambda^3}\frac{1}{e^{hc/\lambda kT}-1}.
\end{equation}
For the case of responsivity weighting, we follow the procedure of \cite{Starkey2017} and \cite{Tie2018} to calculate the multiplicative factor. Specifically, the radiated thermal energy of each annuli of the accretion disk responds to the incident driving variations as
\begin{equation}
\sigma T^4(t) = \sigma T_0^4 \left[1 +  f_c(t-R/c)\right],
\label{eq_temperature}
\end{equation}
where $T_0$ represents the unperturbed temperature of the accretion disk and $f_c$ represents the fraction of incident energy flux thermalized by the disk annuli. For small variations, Equation~(\ref{eq_temperature}) effectively leads to a temperature fluctuation as
\begin{equation}
\frac{\partial T(t)}{\partial f_c} \approx \frac{T}{4}.
\end{equation}
As a result, the multiplicative factor $\chi$ is given by
\begin{equation}
\chi  \equiv \chi_{\rm r} =  \frac{1}{R_{\lambda}} \frac{\int_{R_{\rm in}}^{R_{\rm out}}\frac{\partial B_\lambda(T)}{\partial T}\frac{\partial T}{\partial f_c}R^2 d R}{\int_{R_{\rm in}}^{R_{\rm out}}\frac{\partial B_\lambda(T)}{\partial T}\frac{\partial T}{\partial f_c}R d R},
\label{eq_chir0}
\end{equation}
where the subscript ``r'' represents responsivity weighting.

In Appendix~\ref{sec_app_factor}, we show how to calculate the two multiplicative factors. The two factors depend on the inner and outer disk radius and generally change with wavelength. In extreme cases of $R_{\rm in}\rightarrow0$ and $R_{\rm out}\rightarrow\infty$, the two factors are independent on wavelength and can be expressed analytically with the help of the zeta and gamma functions. Table~\ref{tab_xi} lists values of the two factors with different temperature profiles with $R_{\rm in}\rightarrow0$ and $R_{\rm out}\rightarrow\infty$. For the standard accretion disk model ($\beta=4/3$), we have $\chi_{\rm e}=3.36$ and $\chi_{\rm r}=5.04$, which correspond to $X=2.49$ and $3.36$, respectively.  Table~\ref{tab_xi} also illustrates that the two factors sensitively change with temperature profiles. As $\beta$ increases from 1 to 2, both $\chi_{\rm e}$ and $\chi_{\rm r}$ increase by more than one order of magnitude. 

For finite inner and outer disk radius, $\chi_{\rm e}$ and $\chi_{\rm r}$ can only be calculated numerically. In Figure~\ref{fig_factor}, we plot the two factors with different inner and outer disk radius under the standard disk model ($\beta=4/3$). Both $\chi_{\rm e}$ and $\chi_{\rm r}$ increase with the inner/outer disk radius because of the increasing fraction of radiation from large disk annulus. Moreover, as expected, the inner (outer) disk radius mainly affects the factors at short (long) wavelength. 

It is worth stressing that the disk size measurement through fitting inter-band time delays does not rely on the multiplicative factor. It is only required when comparing the measured disk size against the disk model.

\begin{figure}[t!]
\centering 
\includegraphics[width=0.30\textwidth]{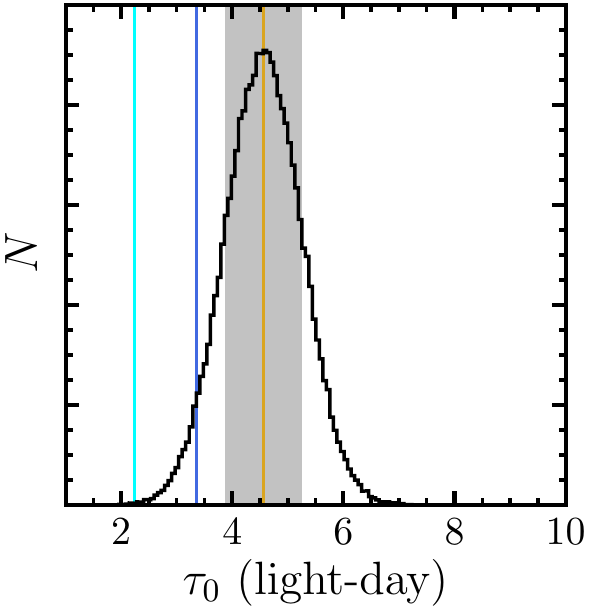}
\caption{The posterior distribution of the disk size $\tau_0$ under Case~I. The shaded area represents the 68.3\% confidence interval and the yellow vertical line represent the best estimate. The blue and cyan lines represent the anticipated disk sizes ($\tau_0=3.35$ and 2.24 light-days) of the standard accretion disk model with $\chi=5.04$ and 3.36, respectively.}
\label{fig_caseI}
\end{figure}

\section{Measuring Accretion Disk Structures}
\label{sec_disk_meausre}
We use Equation~(\ref{eq_tau4}) to fit the observed time delays from ICCF analysis. The present sample size does not allow us to reliably constrain the full parameters ($\tau_0$, $\beta$, $\gamma$, and $\delta$). Therefore, inspired by the energy equation of the generalized accretion disk model (see Equation~\ref{eq_gendisk}), we design the following cases by fixing some of the parameters.
\begin{itemize}
\item[$\bullet$] Case I: fixing $\beta=4/3$, $\gamma=1/3$, and $\delta=1/3$. 

In this case, only $\tau_0$ is free to determine so that we can test the standard accretion disk model by comparing the obtained $\tau_0$ against the anticipated value.

\item[$\bullet$] Case II: fixing $\gamma/\beta=1/4$ and $\delta/\beta=1/4$. 

In this case, $\tau_0$ and $\beta$ are free parameters to determine so that we measure the disk size and temperature profile simultaneously.

\item[$\bullet$] Case III: fixing $\beta=4/3$.

In this case, the disk's temperature profile is the same as in the standard accretion disk model. The rest parameters $\tau_0$, $\gamma$, and $\delta$ are free to determine.
\end{itemize}
Table~\ref{tab_case} summarizes the fixed parameters in the above three cases. We employ a Markov-chain Monte Carlo method implemented with the \texttt{emcee}\footnote{\texttt{emcee} is available at \url{https://github.com/dfm/emcee}.} package to optimize the fitting (\citealt{Foreman-Mackey2013}). We assign the best values by the medians and the uncertainties by the 68.3\% confidence intervals of posterior distributions of the parameters. We adopt a logarithm prior for $\tau_0$ and uniform priors for the rest parameters. The prior range for $\tau_0$ is $(0, 10)$ light-days, for $\beta$ is $(0, 5)$, for $\gamma$ is $(-1, 1)$, and for $\delta$ is $(-1, 1)$. In Table~\ref{tab_case}, we also list the inferred best values and uncertainties of the parameters. Below we present the results for each case.

\begin{figure}[t!]
\centering
\includegraphics[width=0.47\textwidth]{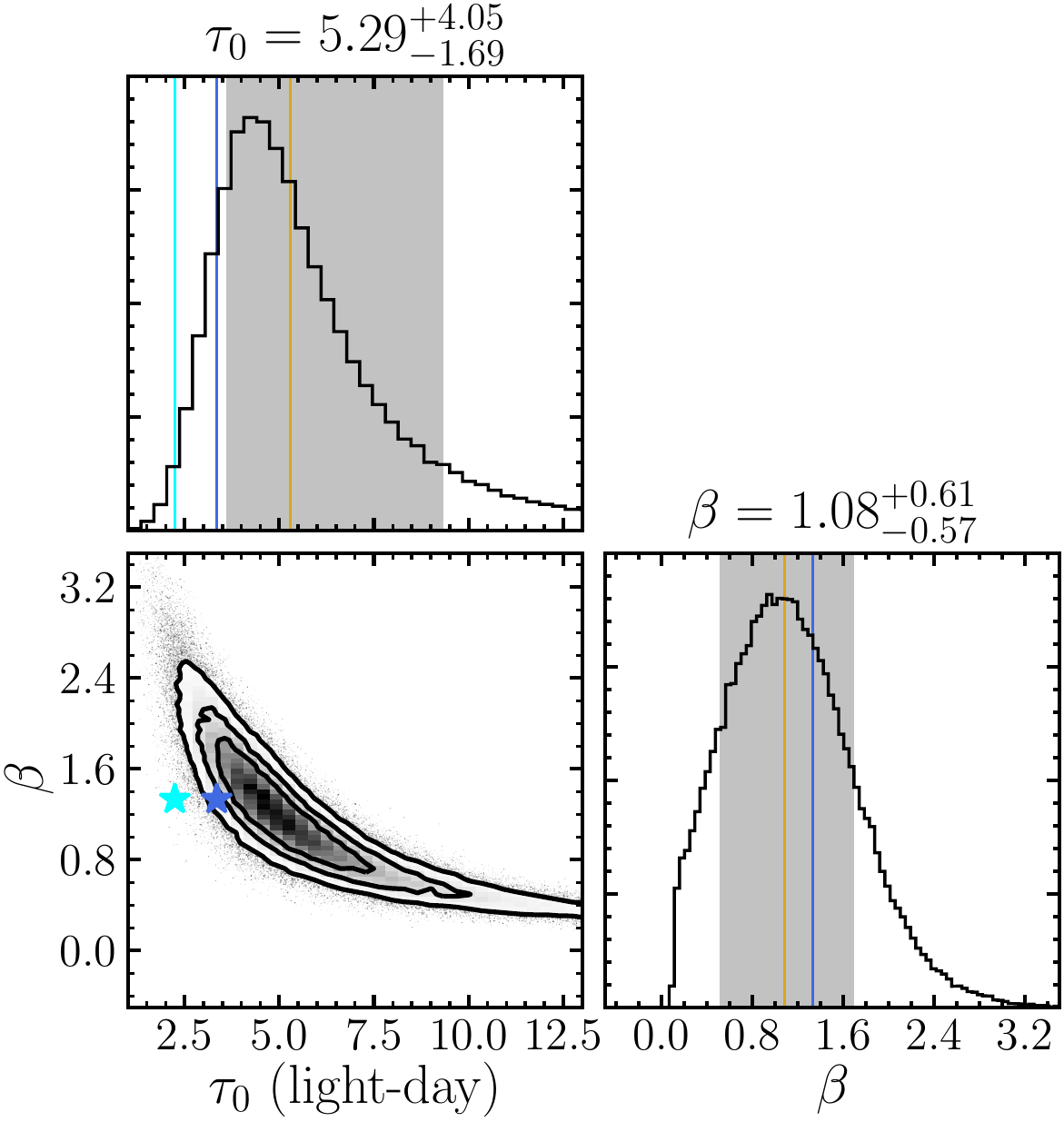}
\caption{
The posterior distributions of the disk size $\tau_0$ and temperature profile parameter $\beta$ for Case II. The orange lines and grey shaded bands show the median values and 68.3\% confidence intervals. For $\tau_0$, the blue and cyan lines (stars) show the theoretical values of the standard accretion disk model with $\chi_{\rm e}=3.36$ and $\chi_{\rm r}=5.04$, respectively. For $\beta$, the blue line (stars) shows the theoretical value of $4/3$. The contours are at 1$\sigma$, 1.5$\sigma$, and 2$\sigma$ levels.}
\label{fig_caseII}
\end{figure}

\begin{figure*}
\centering
\includegraphics[width=0.7\textwidth]{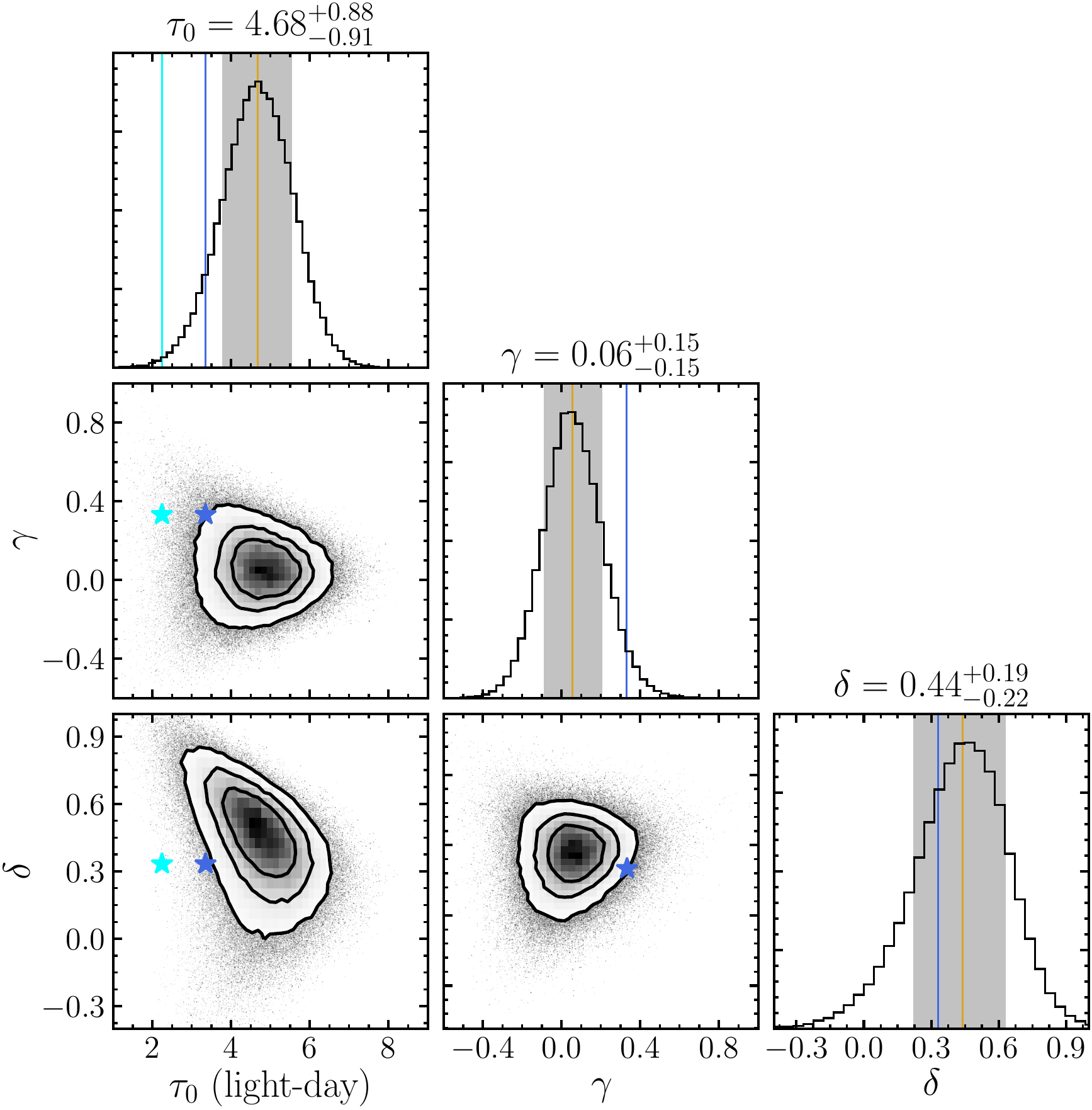}
\caption{
The posterior distributions of $\tau_{0}$, $\gamma$ and $\delta$ for Case III. The orange lines and grey shaded bands show the median values and 68.3\% confidence intervals. For $\tau_0$, the blue and cyan lines (stars) show the theoretic values from the standard accretion disk model with $\chi_{\rm e}=3.36$ and $\chi_{\rm r}=5.04$ , respectively. For other parameters, the blue lines (stars) show the theoretical values. The contours are at 1$\sigma$, 1.5$\sigma$, and 2$\sigma$ levels.
}
\label{fig_caseIII}
\end{figure*}
\subsection{Case I}
In this case, we fix $\beta=4/3$, $\gamma=1/3$, and $\delta=1/3$, and only set $\tau_0$ free. Figure~\ref{fig_caseI} shows the obtained posterior distribution of $\tau_0$. The best estimate is $\tau_{0} = 4.56_{-0.69}^{+0.69}$~light-days. This value is about 1.36 times the prediction of 3.35 light-days with $\chi_{\rm r}=5.04$ at a $2\sigma$ level and 2.06 times the prediction of 2.24 light-days with $\chi_{\rm e}=3.36$ at a $3\sigma$ level. 

\subsection{Case II}
In this case, we fix $\gamma=\beta/4$ and $\delta = \beta/4$ but set $\tau_0$ and $\beta$ free. This means that the radiated flux per unit disk's surface area linearly depends on black hole mass and accretion rate as in the standard accretion disk model (see Equation~\ref{eq_gendisk}), namely,
\begin{equation}
\sigma T^4 \propto M_\bullet \dot M.
\end{equation}

Figure~\ref{fig_caseII} plots the posterior distributions of $\tau_0$ and $\beta$. The disk size parameter $\tau_{0}$ is strongly anti-correlated with the temperature profile parameter $\beta$ as implied from Equation~(\ref{eq_tau4}). Because of the limited redshift range and wavelength coverage of the present sample, there appears a long tail toward the corner of small $\beta$ and large $\tau_0$ in the contour between $\tau_0$ and $\beta$. The best estimate for $\tau_0$ is $5.29_{-1.69}^{+4.05}$~light-days, consistent within uncertainties with the inferred value of Case I.  The best estimate of $\beta=1.08_{-0.57}^{+0.61}$, in agreement with  uncertainties with the fiducial value of $4/3$ for the standard accretion disk model.

\subsection{Case III}
In this case, we fix $\beta=4/3$ to force the temperature profile to follow the same radial dependence ($T\propto R^{-3/4}$) as in the standard accretion disk model. Figure~\ref{fig_caseIII} shows the posterior distributions of $\tau_0$, $\gamma$, and $\beta$. The inferred disk size is $\tau_{0} = 4.68_{-0.91}^{+0.88}$~light-days, again consistent within uncertainties with the measurements of Cases I and II. The parameter of $\delta=0.44_{-0.22}^{+0.19}$ remarkably agrees with the fiducial value of $1/3$ for the standard disk model (see Equation~\ref{eq_tau3}). However, the parameter $\gamma=0.06_{-0.15}^{+0.15}$, lower than the fiducial value of $1/3$ at a 2$\sigma$ level. This discrepancy may arise from the adopted virial factor $f_{\rm BLR}$ that neglects the dependence on black hole mass.

\begin{deluxetable*}{lccccc}
\renewcommand\arraystretch{1.2}
\tablecolumns{5}
\tabletypesize{\footnotesize}
\tabcaption{\centering Three cases of fitting and the inferred values of parameters.  \label{tab_case}}
\tablehead{
\colhead{} &
\colhead{~~~~~~~Case~~~~~~~} &
\colhead{~~~~~~~$\tau_0$ (light-day)~~~~~~~} &
\colhead{~~~~~~~~$\beta$~~~~~~~~~} &
\colhead{~~~~~~~~$\gamma$~~~~~~~~} &
\colhead{~~~~~~~~$\delta$~~~~~~~~}
}
\startdata 
{   }     & Case I   & LogUniform(0, 10) & fixed(4/3) & fixed(1/3) & fixed(1/3)\\
Prior     & Case II  & LogUniform(0, 10) & Uniform(0, 5) & fixed($\beta/4$) & fixed($\beta/4$)\\
{   }     & Case III & LogUniform(0, 10) & 4/3 & Uniform(-1, 1) & Uniform(-1, 1)\\\hline
{   }     & Case I   &  $4.56_{-0.69}^{+0.69}$    & \nodata & \nodata & \nodata \\
Posterior & Case II  &  $5.29_{-1.69}^{+4.05}$    & $1.08_{-0.57}^{+0.61}$ & \nodata & \nodata \\
{   }     & Case III &  $4.68_{-0.91}^{+0.88}$    & \nodata                & $0.06_{-0.15}^{+0.15}$ & $0.44_{-0.22}^{+0.19}$ \\
\enddata
\end{deluxetable*}

\begin{deluxetable*}{ccccccc}
\renewcommand\arraystretch{1.0}
\tablecolumns{5}
\tabletypesize{\footnotesize}
\tabcaption{\centering A summary of disk size measurements.  \label{tab_compare}}
\tablehead{
\colhead{Data} &
\colhead{Sample Size} &
\colhead{Median Cadence} &
\colhead{Duration} &
\colhead{Bands} &
\colhead{RM } &
\colhead{Reference}
}
\startdata 
{Pan-STARRS}  & 39   & 3 days & 3.3 years & $g,r,i,z$ & No &\cite{Jiang2017}\\
{OzDES}       & 15   & 7 days & 180 days  & $g,r,i,z$ & No &\cite{Mudd2018}\\
{SDSS-RM}     & 95   & 4 days & 180 days  & $g,i$ & Yes    &\cite{Homayouni2019}\\
{OzDES}       & 23   & 2 days & 3.3 years & $g,r,i,z$ & No &\cite{Yu2020}\\
{ZTF}         & 19   & 3 days & 3 years  &  $g,r,i$  & Yes    &\cite{Jha2021}\\
{ZTF}         & 92   & 3 days & 200 days  & $g,r,i$ & No  & This Work\\
\enddata
\tablecomments{``RM'' means reverberation mapping of broad emission lines, from which black hole mass can be estimated, instead of using the $R_{\rm{H}\beta}$-$L_{5100}$ relation.}
\end{deluxetable*}

\begin{figure*}[th!]
\centering
\includegraphics[width=0.95\textwidth]{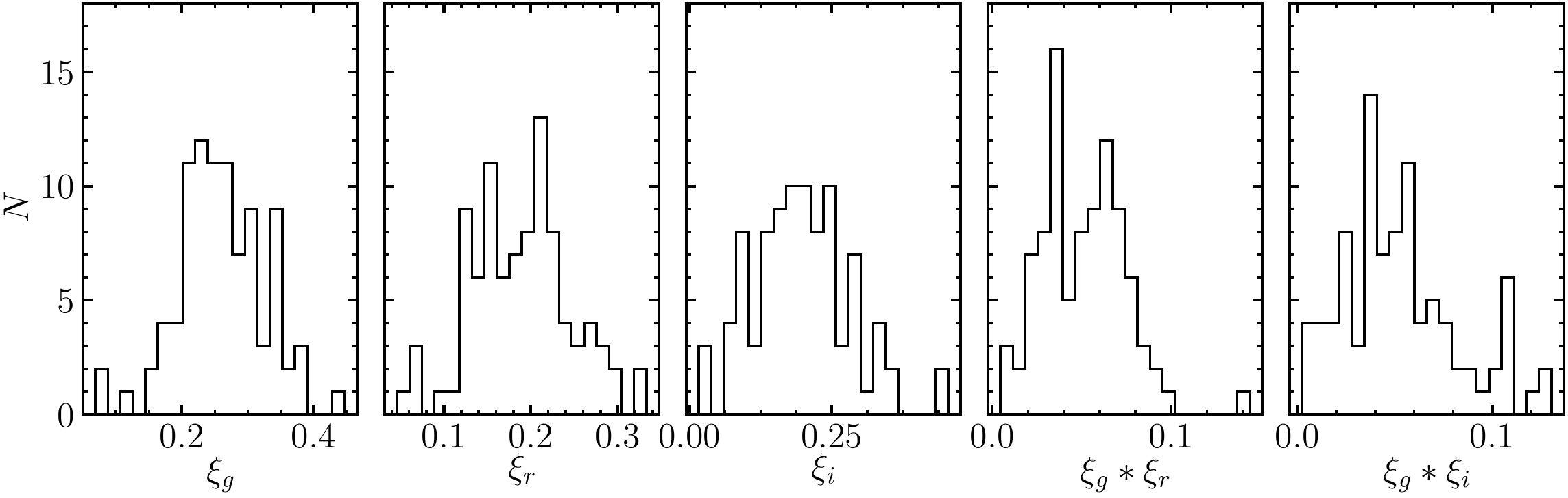}
\caption{ The three left panels are the flux fractions of broad emission lines  in $g$, $r$ and $i$ bands respectively. The rightmost two panels are products of the flux fractions of broad emission lines in $gr$ and $gi$ bands. }
\vspace{+0.5cm}
\label{fig_blr}
\end{figure*}

\begin{figure*}[th!]
    \centering 
    \includegraphics[width=0.7\textwidth]{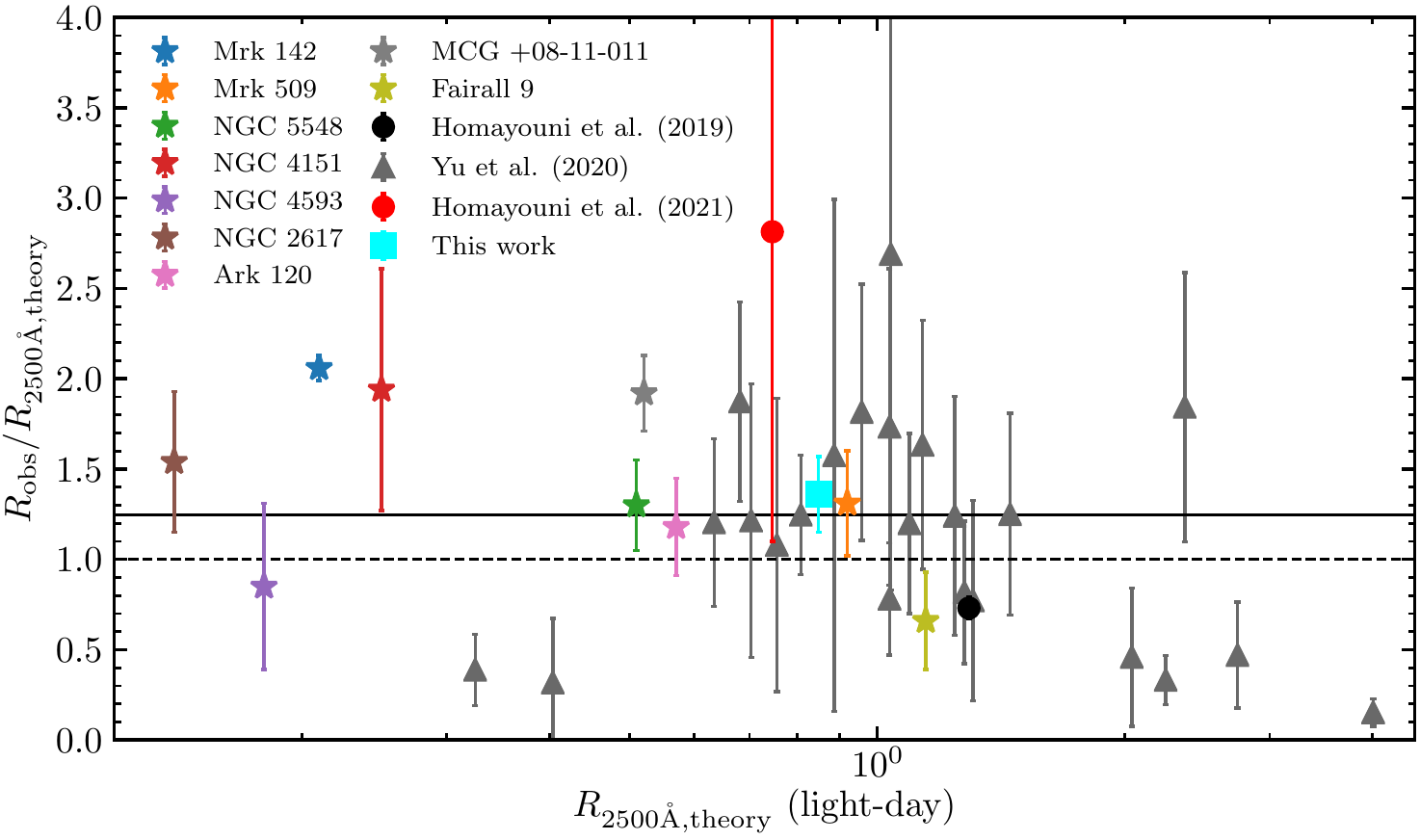}
    \caption{The ratio of the measured disk size to theoretical size at 2500~{\AA} based on the standard accretion disk model with the multiplicative factor $\chi=5.04$. The solid line represent the median ratio 1.24 and the dashed line represent the ratio equal to one. }
    \label{compare}
\end{figure*}

\subsection{Comparisons with Previous Studies}
There were several previous studies that measured accretion disk sizes using AGN samples from time-domain surveys (see Table~\ref{tab_compare} for a brief summary). \cite{Homayouni2019} analyzed continuum time delays between $i$ and $g$ bands of 95 quasars in the Sloan Digital Sky Survey Reverberation Mapping project and did not  find evidence that the derived disk size deviates from the theoretical expectation of the standard accretion disk model. \cite{Mudd2018} and \cite{Yu2020} measured disk sizes of 15 and 22 quasars using $griz$ light curves selected from different fields of the Dark Energy Survey, respectively. They both drew a similar conclusion as \cite{Homayouni2019}. However, \cite{Homayouni2021} used UV-optical continuum reverberation mapping observations of eight quasars and found that the disk size is 2-4 times the expectation of the standard accretion disk model if assuming $\chi=3.36$. \cite{Jiang2017} performed disk size measurements of 39 AGNs based on the $griz$ light curves from the Pan-STARRS survey and also concluded that the disk sizes were bigger than expected by a factor of 2-3 with $\chi=3.36$. Recently, \cite{Jha2021} cross matched the previously reverberation-mapped AGNs (using the broad H$\beta$ line) with the ZTF archival data and compiled a sample of 19 AGNs. Their obtained disk sizes are also found to be larger than expected (with $\chi=3.36$). 

The high-precision measurements from individual nearby AGNs also generally indicated a disk size larger by a factor of 2-3 than the theoretical expectation (e.g., \citealt{Fausnaugh2015,Cackett2017,Edelson2019}). However, an exception is Fairall 9, which shows a delay spectrum following the standard accretion disk model  (\citealt{Hern2020}) in both amplitude and power index. These studies generally adopted $\chi=3.36$. 

As a comparison, our results in all three cases show that regardless of emissivity weighting ($\chi=3.36$) or responsivity weighting ($\chi=5.04$), the disk size is larger than the theoretical expectation. In Figure~\ref{compare}, we plot the ratio of the inferred to the theoretical disk sizes from previous studies and this work. For the sake of consistency, we uniformly use the same factor of responsivity weighting, namely, $\chi=5.04$. In addition, different studies reported disk sizes at different reference wavelengths. We uniformly convert the reference wavelength to 2500~{\AA}  using a relation $\tau_{0, 2500\text{\AA}} / \tau_{0, \lambda_0}= (\lambda_0 /2500 \text{\AA})^{-4/3}$ based on the standard accretion disk model. Figure~\ref{compare} implies that if regardless of the large scatters, the averaged disk size is still larger than the theoretical size. The median ratio of the measured to theoretical disk sizes is 1.24.

As for the temperature profile, our results agree with previous studies (e.g., \citealt{Homayouni2019,Cackett2017,Edelson2019}), which overall found $\tau \propto \lambda^{4/3}$ as expected from the standard accretion disk model, although  our results have relatively large uncertainties due to the limited sample size and filters. 

\begin{figure*}[t!]
\centering
\includegraphics[width=0.55\textwidth]{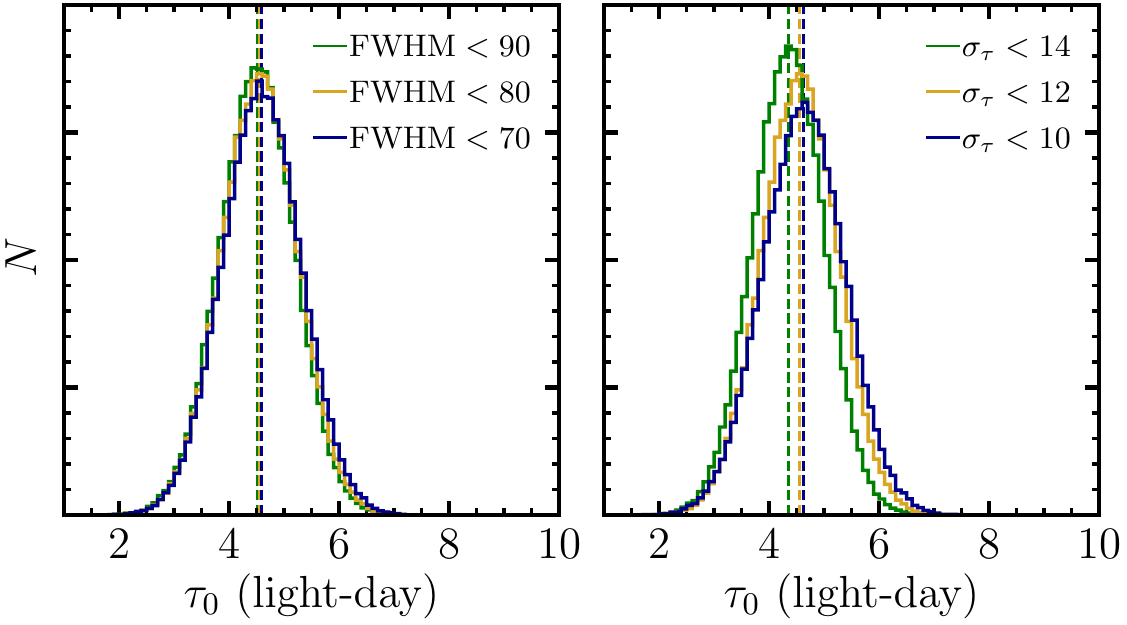}
\caption{ The distributions of the inferred disk size $\tau_0$ under case I for different 
selection criteria. (Left) the ACF FWHMs of mock light curves are restricted to $<$90, 80, and 70 days. (Right) the uncertainties of time delays of mock light curves are restricted to $\sigma_\tau<$14, 12, and 10 days. 
}
\label{fig_select}
\end{figure*}

\section{Discussions} 
\label{sec_discussion}
\subsection{The Influences of Broad Emission Lines}\label{sec_dis_bel}
In Section~\ref{sec_test}, we assess the influences of broad emission lines through Monte Carlo simulations and select out the final sample with the minimum influences. Here we make a more thorough discussion. In the presence of broad emission lines, the light curve in each band has an additional component, i.e., $f_{X}=f^{c}_{X}+f^{l}_{X}$, where the superscript ``$c$'' denotes the intrinsic continuum flux and ``$l$'' denotes the line flux. The cross-correlation function (CCF) between light curves of two bands, say, $X$ and $Y$, is then given by 

\begin{eqnarray}
{\rm CCF}(f_X, f_Y) &=& \frac{E\left[(f_X-\bar{f}_X)(f_Y-\bar{f}_Y)\right]}{\sigma\left[f_X\right]\sigma\left[f_Y\right]}\nonumber\\
&=&(1-\xi_X)(1-\xi_Y) {\rm CCF}(f_X^c, f_Y^c) \nonumber\\
&+&\xi_X(1-\xi_Y){\rm CCF}(f_X^l, f_Y^c) \nonumber\\
&+& (1-\xi_Y)\xi_Y{\rm CCF}(f_X^c, f_Y^l)\nonumber\\
&+& \xi_X\xi_Y{\rm CCF}(f_X^l, f_Y^l),
\label{eq_ccf}
\end{eqnarray}%
where $E$ and $\sigma$ represent the expectation and standard deviation of the light curves, and $\xi_X=f_X^l/(f_X^c+f_X^l)$ and  $\xi_Y=f_Y^l/(f_Y^c+f_Y^l)$ represent the flux fractions in $X$ and $Y$ bands contributed from broad emission lines, respectively. 
 For simplicity, we assume a common characteristic time delay of $\tau_{\rm BLR}$ for all broad emission lines. In the far right-hand side of Equation~(\ref{eq_ccf}), the second and third terms give rise to CCF peaks at about $-\tau_{\rm BLR}$ and $+\tau_{\rm BLR}$, respectively. The fourth term yields a CCF peak at zero time delay. 
 It is clear that the influences of these terms to the final time delay depend on $\xi_X$, $\xi_Y$ and their product $\xi_X*\xi_Y$. 
\begin{itemize}
 \item If $\xi_X > \xi_Y$, the second term will be dominated over the other two terms and therefore the resulting time delay 
 will tend to be shortened (compared with that without broad emission lines).
 \item If $\xi_X < \xi_Y$, we have the opposite case and the time delay 
 will tend to be enlarged.
 \item If $\xi_X\rightarrow 1$ and $\xi_Y\rightarrow 1$, the forth term will be dominated and the time delay tends to be zero.
\end{itemize}

In Figure~\ref{fig_blr}, we show the flux fractions of broad emission lines in $g$, $r$, and $i$ bands,
together with their products. We find that generally $\xi_g>\xi_r$ and $\xi_g > \xi_i$, implying that 
the presence of broad emission lines tends to shorten the time delays.
In Appendix~{\ref{app_sim}, we perform simulations to illustrate how time delays are 
quantitatively affected for different flux fractions of broad emission lines, which is in agreement with the above simple 
arguments. }

\begin{figure*}
\centering
\includegraphics[width=0.85\textwidth]{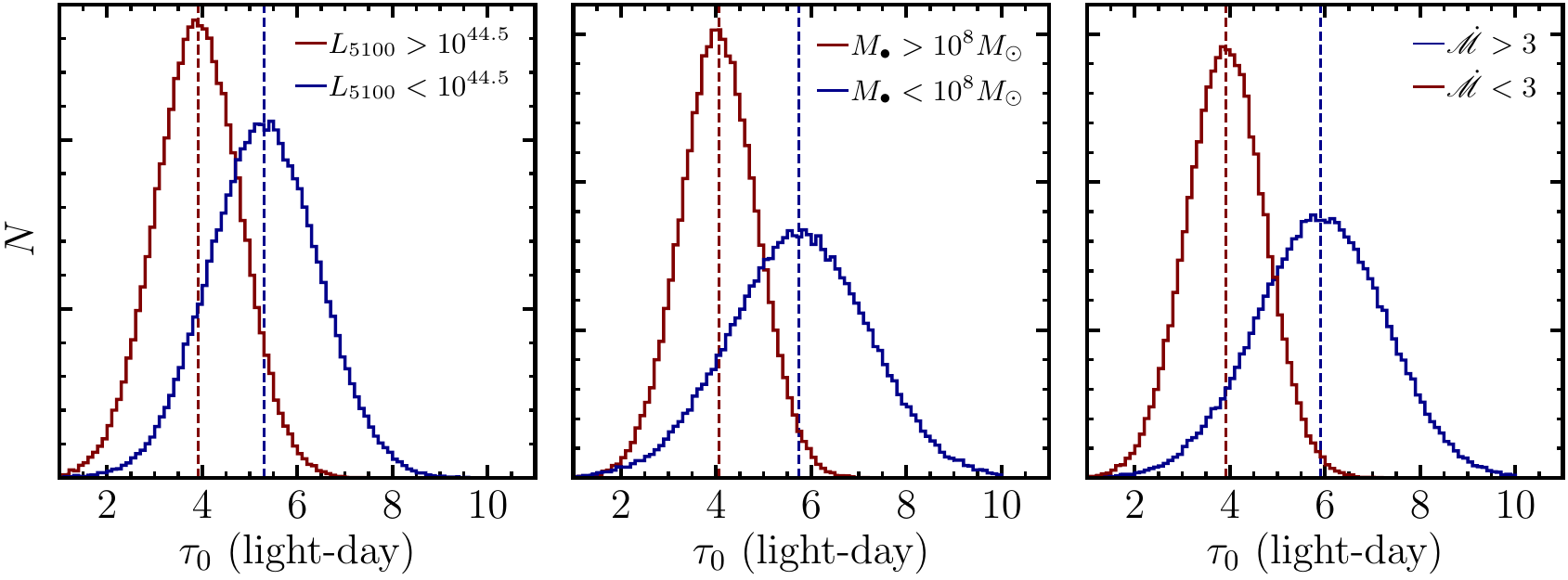}
\caption{The posterior distributions of the disk size $\tau_0$ under Case I for subsamples of (left) 5100~{\AA} luminosity $L_{5100}<10^{44.5}\rm{~erg\ s^{-1}}$ and  $>10^{44.5}\rm{~erg\ s^{-1}}$, (middle) black hole mass $M_{\bullet}>10^{8}M_{\odot}$ and  $<10^{8}M_{\odot}$,
and (right) Eddington ratio $\mathscr{\dot M}<3$ and $>3$.
}
\label{fig_mdot}
\end{figure*}

\subsection{The Selection Criteria}
To test the influences of the selection criteria in Section~\ref{sec_criteria}, we adjust the restrictions  on the ACF FWHM and mock uncertainties and repeat the above analysis procedures for Case I. In the left panel of Figure \ref{fig_select}, we restrict the ACF FWHM to be smaller than 90, 80, and 70 days and obtain the disk size $\tau_0=4.51^{+0.68}_{-0.67}$, $4.56^{+0.69}_{-0.69}$, and $4.59^{+0.72}_{-0.72}$ days, respectively. As the restriction  of the ACF FWHM decreases, the inferred disk size is almost unchanged, indicating that it is insensitive to the limit of the ACF FWHM. Meanwhile, the AGN sample size is reduced from 96 to 87, but the quality of the inter-band time delays improve in terms of errors. These two factors lead to almost similar uncertainties of the inferred disk sizes for different criteria of the ACF FWHM. This also plausibly implies that the bigger-than-expected disk size is true and not caused by measurement errors. 

In the right panel of Figure \ref{fig_select}, we set an upper limit of the mock uncertainty of the simulated light curves $\sigma_\tau<$10, 12, and 14 days (see Section~\ref{sec_test}). The correspondingly disk sizes are $\tau_0=4.62^{+0.74}_{-0.75}$, $4.56^{+0.69}_{-0.69}$, and $4.35^{+0.65}_{-0.65}$ days, respectively. The inferred disk size gets slightly larger as $\sigma_\tau$ decreases (but is consistent within uncertainties). Meanwhile, the associated uncertainty slightly increases because the AGN number decreases (from 110 to 75). A lower limit of $\sigma_\tau$ means that AGNs with relatively higher-quality time delays will be selected. In this sense, our main result that the measured disk size is bigger than expected is robust. 
 
\subsection{The Lamp-post Model}
The lamp-post model has  an additional characteristic factor $\kappa$, 
which describes the ratio of external to local viscous heating. 
The disk size in the lamp-most model is given by (e.g., \citealt{Cackett2007})
\begin{equation}
\tau_{\rm d}=\left[\frac{15 G C_{bol}(3+\kappa)}{16 \eta h \pi^{6} c^{4} }\right]^{1/3} \lambda_{0}^{4/3}X^{4/3}M_{\bullet}^{1/3} \ L_{5100}^{1/3},
\end{equation}
where $\kappa=2(1-A)L_{X}H/GM_{\bullet}\dot{M}$, $L_X$ is the X-ray luminosity, $H$ is the height of the X-ray lamp above the disk, and $A$ is the albedo of the disk. Generally, the value of $\kappa$ is on the order of unity, which means that X-ray and viscous heating contribute equal amounts of energy to the disk radiation. Therefore, we expect that compared to the standard accretion disk model, the lamp-post model does not significantly change the theoretically expected disk size.

\subsection{Dependence on AGN Properties}
We divide the sample into subsamples according to the 5100~{\AA} luminosity $L_{5100}$, black hole mass $M_\odot$, and accretion rate $\mathscr{\dot M}$, respectively. We then repeat the above fitting analysis of Case I for each subsample and show the posterior distributions of $\tau_0$ in Figure~\ref{fig_mdot}. In the left panel, there are 46  AGNs with $L_{5100}<10^{44.5}~\rm erg~s^{-1}$ and 46 AGNs with $L_{5100}>10^{44.5}~\rm erg~s^{-1}$. The corresponding best estimated disk sizes are $\tau_0=3.91^{+0.88}_{-0.88}$ days and $5.29^{+1.14}_{-1.14}$ days, respectively. In the middle panel, we obtain the disk size $\tau_0=5.74^{+1.45}_{-1.45}$ days and $4.05^{+0.81}_{-0.80}$ days for the black hole mass $M_\bullet<10^8M_\odot$ (45 AGNs) and $>10^8M_\odot$ (47 AGNs). The right panel of Figure~\ref{fig_mdot} compares the disk sizes for sub-Eddington ($\mathscr{\dot M}<3$) and super-Eddington ($\mathscr{\dot M}>3$) subsamples, which consist of 56 and 36 AGNs, respectively. The best estimates are $\tau_0=3.91^{+0.82}_{-0.82}$ days for sub-Eddington subsample and $5.90^{+1.35}_{-1.35}$ days for super-Eddington subsample. The overall large uncertainties mean that the differences between the subsamples are not significant and thereby further tests with future ZTF data releases are warranted to draw solid conclusions. However, it is worth mentioning that the recent study of \cite{Sun2021} showed that low-luminosity AGNs favor large disk sizes by surveying available disk size measurements from the literature (see Figure 1 therein). This is remarkably consistent with our results in the left panel of Figure~\ref{fig_mdot}. 
 
From the theoretical point of view, we normalize the disk size to a fixed black hole mass and luminosity (thus to a fixed accretion rate; see Equations~\ref{eq_size1} and \ref{eq_size2}), therefore, we expect that the disk size $\tau_0$ is independent on AGN properties. However, there are two possibilities that may cause the observed dependence in Figure~\ref{fig_mdot}. First, we adopt a constant virial factor $f_{\rm BLR}$ when estimating the black hole mass using Equation~(\ref{eq_mass}). Dynamical modeling of broad-line regions generally showed that $f_{\rm BLR}$ might depend on AGN properties and change from object to object (e.g., \citealt{Pancoast2011, Pancoast2014, Li2013, Li2018, Grier2017, Williams2018}). Observational calibrations of the virial factor 
also tend to support this conclusion (e.g., \citealt{Ho2014, Meja-Restrepo2018, Yu2019}). If there are systematic correlations between $f_{\rm BLR}$ and luminosity $L_{5100}$, black hole mass $M_\bullet$, or accretion rate $\mathscr{\dot M}$, using a constant $f_{\rm BLR}$ will lead to the apparent dependence of $\tau_0$ on AGN properties. Such a bias can be eliminated once we have a solid understanding of the virial factor in the future. Second, it is possible that the disk's temperature profile does change with the accretion rate. For example, the standard and slim disk models predict different temperature profiles\footnote{However, we note that the temperature profiles of slim disks ($T\propto R^{-1/2}$) become different from those of standard disks ($T\propto R^{-3/4}$) only within the photon-trapping radius (e.g., \citealt{Wang1999b}).} (\citealt{Shakura1973, Abramowicz1988}). There are also other physical processes that may contribute to the correlation between $\beta$ and $\mathscr{\dot M}$ (e.g., see \citealt{Sun2021} and \citealt{Kammoun2021}). Figure~\ref{fig_caseII} implies that $\tau_0$ and $\beta$ are highly anti-correlated. As a result, a shallower temperature profile (larger $\beta$) will lead to a smaller disk size $\tau_0$ and vice versa. In our sample, we find that high-luminosity and high-mass black holes generally have low accretion rates, giving rise to the dependence of $\tau_0$ on luminosity and black hole mass.

\subsection{The Virial Factor}
The value of the virial factor affects black hole mass estimation and thereby the expected disk size (based on the standard disk model) as
\begin{equation}
\tau_{\rm d}\propto M_\bullet^{1/3}\propto f_{\rm BLR}^{1/3}.
\label{eq}
\end{equation}
In our calculations, we adopt $f_{\rm BLR}=1.12\pm 0.31$ from Woo et al. (2015). With this value, we find that the measured disk size is 
1.36 and 2.06 times the theoretical expectation using a responsivity-weighted and emissivity-weighted factor, respectively.
If we manually align the measured disk size with the theoretical expectation, the virial factor has to be increased to $f_{\rm BLR}=2.82$ and $9.80$, respectively. For a disk-like BLR with an opening angle $\theta$ and inclination of $i$, the virial factor (in terms of FWHM) can be expressed as (e.g., \citealt{Collin2006})
\begin{equation}
f_{\rm BLR} \approx 0.25(\sin^2\theta + \sin^2 i)^{-1},
\end{equation}
where the coefficient $0.25$ is owling to using FWHM as the measure of line widths.
To produce a virial factor of $f_{\rm BLR}=2.82$ and $9.80$, it is easy to gather that both the inclination and opening angles are needed to be small (say, $\lesssim15^\circ$). This seems unreasonable in consideration of the expected diverse distribution of inclination. Also, previous BLR reverberation mapping showed evidence for a thick BLR disk, namely, a large opening angle (e.g., \citealt{Peterson2014}).

\subsection{The Internal Extinction of AGNs}
It is clear that the AGN extinction leads to underestimation of the AGN luminosities and thereby the theoretical accretion disk size. We use the H$\alpha$/H$\beta$ ratio to estimate the AGN internal extinction.
We limited the redshift to $z<0.35$ so as to have both H$\alpha$ and H$\beta$ lines in the spectral wavelength coverage. There are 49 AGNs
in our sample satisfying this criterion. The obtained H$\alpha$/H$\beta$ ratio range from 2.6 to 5.1. The median ratio is 3.55, generally consistent with that of the SDSS DR7 AGN sample (\citealt{Baron2016}).

If presuming the intrinsic H$\alpha$/H$\beta$ ratio to be 2.72 as in Gaskell (2017), the median ratio of 3.55 corresponds to extinction of $E(B-V)= 0.21$ using the gray dust
model of \cite{Gaskell2017}. This leads to the intrinsic luminosity at 5100~{\AA} larger by a factor of 0.9-2.0  with $R_{V}=3.1-5.5$ (\citealt{Gaskell2007}). As a result, the estimated black hole mass will be larger by 10\%-18\% (Equation~\ref{eq_mass}). The theoretical disk size will be increased to 3.05-3.80 days with the emissivity-weighted factor and 4.57-5.68 days with the responsivity-weighted factor. The latter case can explain the measured disk sizes while in the former case the theoretical disk size is still a bit smaller. As a comparison, if using the extinction curve of the Milky Way (\citealt{Pei1992}), the resulting extinction is $E(B-V)= 0.16$. The intrinsic luminosity at 5100~{\AA} will be larger by a factor of 0.6-1.2 with $R_{V}=3.1-5.5$. With the emissivity-weighted (responsivity-weighted) factor, the theoretical disk size is changed to 2.84-3.20 (3.19-4.80) light-days. 

We note that there are still debates on quantitatively estimating the AGN extinction (e.g., \citealt{McKee1974,Netzer1979,Zotti1985,Gaskell2007,Ferland2013}).
It seems that the recipe of Gaskell (2017) gives a relatively large extinction, which can be regarded as an upper conservative limit.

\section{Conclusions}
\label{sec_conclusion}
We compile a sample of 92 AGNs with $gri$ photometric light curves from the ZTF DR3 archival data and measure their inter-band time delays with the ICCF method. We also use the MICA method (\citealt{Li2016}) and von Neumann estimator (\citealt{chelouche2017}) to cross check the ICCF delays  and find the general consistency (see Appendix~\ref{sec_mica}). We then employ a generalized accretion disk model to fit these time delays with a Markov-chain Monte Carlo technique and put constraints on the disk temperature profile, characteristic disk size, and its dependence on black hole mass and accretion rate. The present sample does not allow us to simultaneously determine the full parameters (the disk size parameter $\tau_0$, the temperature profile parameter $\beta$, and the parameters $\gamma$ and $\delta$ for the dependence on black hole mass and luminosity; see Equation~\ref{eq_tau4}). We design three cases by fixing some parameters according to the energy equation of the generalized disk model (see Equation~\ref{eq_gendisk}). Below by saying the disk size, we refer to the size normalized to a black hole mass of $10^8M_\odot$ and luminosity of $10^{44.5}~\rm erg~s^{-1}$ (see Equation~\ref{eq_tau4}). Our main results are summarized as follows.
\begin{itemize}
    \item The best estimated disk size is $\tau_0=4.56_{-0.69}^{+0.69}$ light-days. This is 1.36 times as large as the theoretical expectation $3.35$ light-days of the standard accretion disk model (namely, steady-state, optically thick, geometrically thin, and with a Keplerian rotation) at a $2\sigma$ level if using the responsivity-weighted factor $\chi=5.04$ and 2.06 times the theoretic predication $2.24$ light-days at a 3$\sigma$ level if using the emissivity-weighted factor $\chi=3.36$.
    
    \item The disk temperature profile parameter $\beta=1.08^{+0.61}_{-0.57}$, consistent within uncertainties with the value of $4/3$ for the standard accretion disk model.
    
    \item  The dependence of  inter-band time delays on luminosity has a power index of $\delta=0.44^{+0.19}_{-0.22}$, agreeing with the theoretical expectation of $1/3$. However, the dependence on black hole mass has a power index of $\gamma=0.06^{+0.15}_{-0.15}$, deviating from the expected value of $1/3$. This may be due to assuming a constant virial factor $f_{\rm BLR}$ and its possible dependence on black hole mass is neglected.
    
    \item Our results tentatively show that the (normalized) disk size might depend on black hole mass, luminosity, and accretion rate (see Figure~\ref{fig_mdot}). However, considering the large uncertainties, this warrants future tests with more high-quality data for further confirmation. 
\end{itemize}

These results are broadly in agreement with previous studies based on large AGN samples (e.g., \citealt{Jiang2017,Homayouni2021, Jha2021}) or individual AGNs with intensive monitoring (e.g., \citealt{Fausnaugh2015, Cackett2017, Edelson2019}). Nevertheless, we remark that 1)  inclusion of the intrinsic AGN extinction may partially reconcile the discrepancy of accretion disk sizes, although there is no consensus on quantitatively estimating the extinction; and 2) limited by the sample size, our results bear larger uncertainties. With increasing data accumulation of the ZTF, future larger AGN samples will place  more restricted constraints on the disk properties. 


\software{\texttt{DASpec} (https://github.com/PuDu-Astro/DASpec),
\texttt{MICA} (\citealt{Li2016}), \texttt{CDNest} (\citealt{Li2020cdnest}), 
\texttt{emcee} (\citealt{Foreman-Mackey2013})}

\section*{acknowledgements}
We thank the referee for useful comments that significantly  improve the manuscript.
We thank P. Du and C. Hu for valuable advice. We acknowledge financial support from the National Natural Science Foundation of China (11833008 and 11991054), from the National Key Research and Development Program of China (2016YFA0400701), and from  China Manned Space Project (CMS-CSST-2021-A06 and CMS-CSST-2021-B11). Y.-R.L. acknowledges financial support from the National Natural Science Foundation of China through Grant No. 11922304, from the Strategic Priority Research Program of CAS through Grant No. XDB23000000, and from the Youth Innovation Promotion Association CAS. L.C.H. acknowledges financial support from the National Science Foundation of China (11721303 and 11991052) and the National Key Research and Development Program of China (2016YFA0400702).

The Zwicky Transient Facility Collaboration is supported by U.S. National Science Foundation through the Mid-Scale Innovations Program (MSIP). This research has made use of the NASA/IPAC Extragalactic Database (NED) which is operated by the Jet Propulsion Laboratory, California Institute of Technology, under contract with the National Aeronautics and Space.

\appendix
\section{Simulation Tests on the Influences of Broad Emission Lines and Data Sampling} 
\label{app_sim}
\begin{figure}[t!]
\centering 
\includegraphics[scale=0.45]{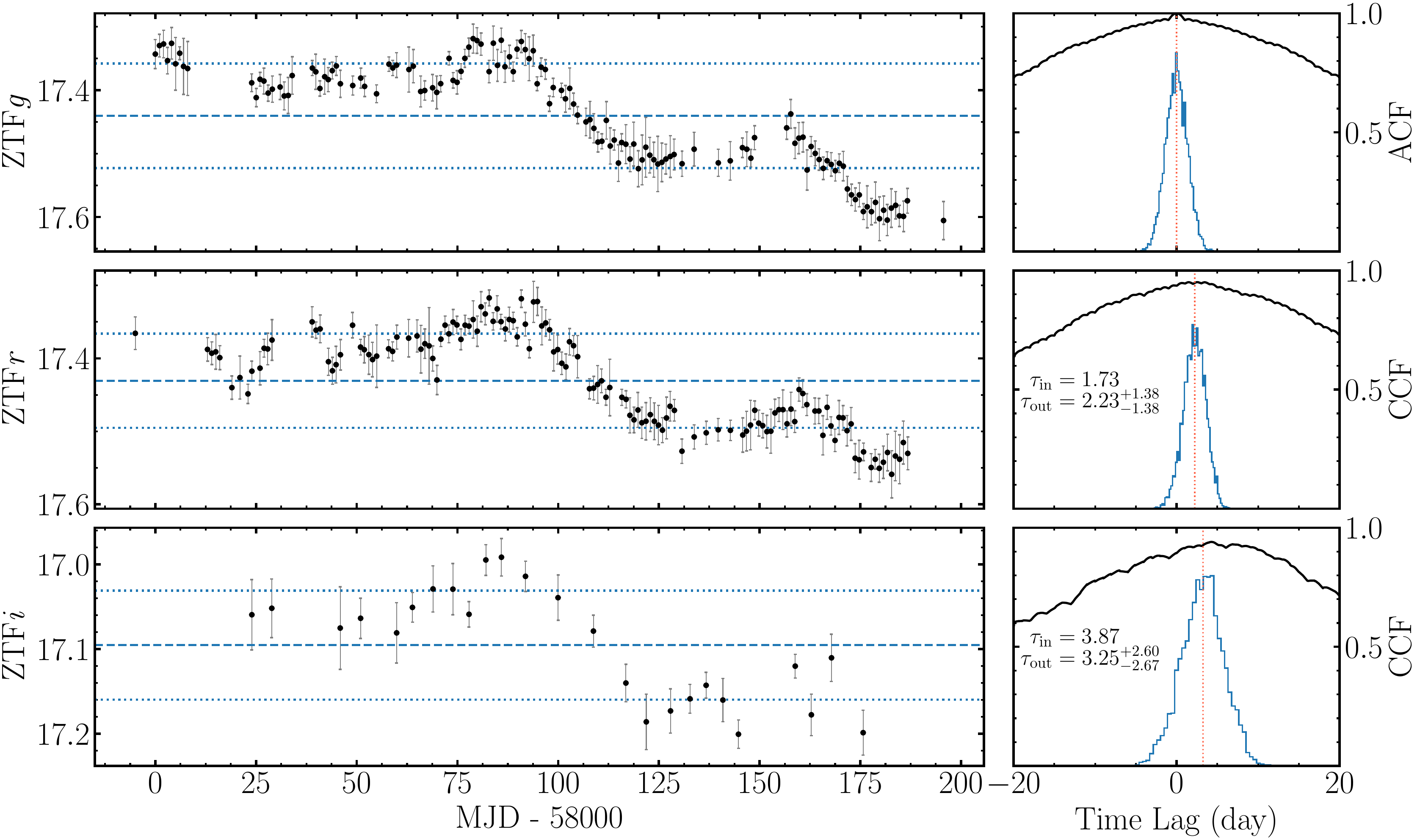}
\caption{
(Left) an example of mock $g$-, $r$- and $i$-band light curves generated according to the observed data in Figure~\ref{ccf}. (Right) from top to bottom panels are the ACF of the $g$-band mock light curve and the ICCFs of $r$- and $i$-band mock light curves with respect to $g$-band one, respectively. The histograms represent the cross-correlation centroid distributions. The input and output time delays  (in units of days) for $r$- and $i$-band are also marked in each panel.
}
\label{sim}
\end{figure}

In Section~\ref{sec_test}, we perform Monte Carlo simulations to test the influences of broad emission lines and data sampling on the observed time delays and select those AGNs with minimized influences.
Here, for the sake of illustration, we show an example of mock light curves and ICCF analysis in Figure~\ref{sim}. The mock light curves are generated according to the observed data in Figure~\ref{ccf}.

\begin{figure}[t]
\centering 
\includegraphics[width=0.6\textwidth]{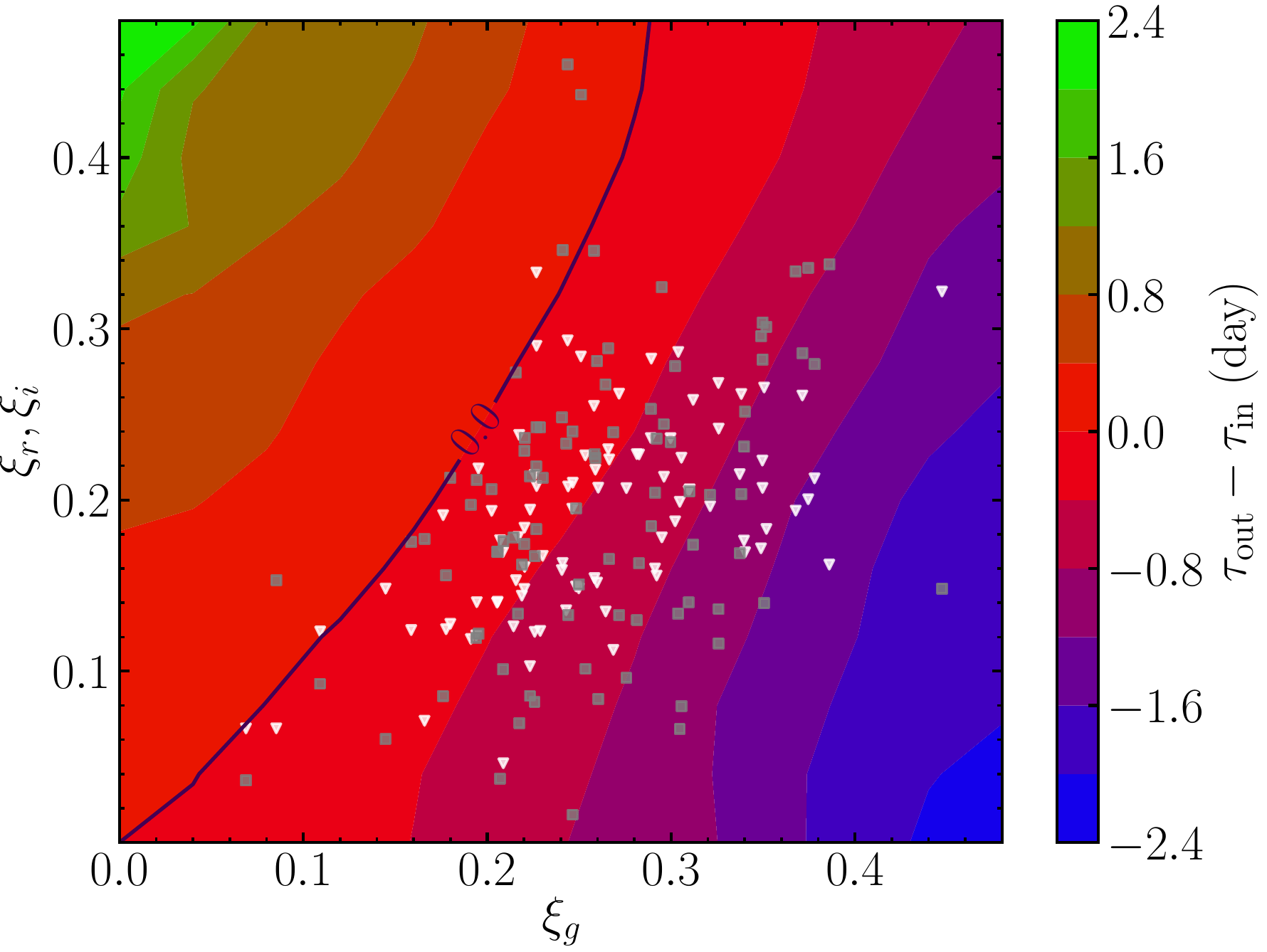}
\caption{ Differences between output and input time delays of two (arbitrary) bands using simulated light curves for different flux fraction of broad emission lines (see Appendix~\ref{app_sim} for a detail). Superimposed are the observed flux fractions of broad emission lines in $g$, $r$, and $i$ bands for our AGN sample. White squares and gray triangles represent $r$ and $i$ bands, respectively.
}
\label{fig_ratio}
\end{figure}

 In addition, we perform simulations to quantitatively illustrate how broad emission lines affect the time delay measurements.
Without loss of generality, we adopt a characteristic time delay of 2 days for between two bands (say, e.g., $g$ and $r$ bands or $g$ and $i$ bands) and 65 days for broad emission lines according to the median black hole mass and 5100~{\AA} luminosity of our sample. Given a pair of flux fractions of broad emission lines in two bands, we use the same procedure described in Section~\ref{sec_test} to generate 1000 sets of light curves. The sampling rate is overall set to 3 days apart. We then use the ICCF method to calculate time delays and derive the mean time delay for the simulated light curves. In Figure~\ref{fig_ratio}, we plot the differences between output ($\tau_{\rm out}$) and input ($\tau_{\rm in}$) time delays for different flux fractions of broad emissions lines.
We also superimpose the observed fractions of broad emission lines in $g$, $r$, and $i$ bands for our sample. As can be seen, the majority of AGNs in our sample are located within the region of $\tau_{\rm out}-\tau_{\rm in}<0$, indicating that the presence of broad emission lines tends to shorten the measured time delays than the realistic values.

\section{Calculating the Multiplicative Factors}
\label{sec_app_factor}
In this section, we show how to calculate the multiplicative factors defined in Section \ref{sec4.2}. We first introduce a new variable $x=hc/kT\lambda$. With a temperature profile $T\propto R^{-1/\beta}$,  we have
\begin{equation}
R\propto x^\beta~{\rm and} ~d R \propto x^{\beta-1}dx.
\end{equation}
As a result, the emissivity-weighted factor can be recasted as
\begin{eqnarray}
 \chi_{\rm e} =
\frac{1}{R_{\lambda}}\frac{\int_{R_{\rm in}}^{R_{\rm out}}B_\lambda(T)R^2 d R}{\int_{R_{\rm in}}^{R_{\rm out}}B_\lambda(T)R d R}
=\frac{\int_{x_{\rm in}}^{x_{\rm out}} \frac{1}{e^{x}-1 }x^{3\beta-1}dx} {\int_{x_{\rm in}}^{x_{\rm out}}\frac{1}{e^{x}-1}x^{2\beta-1} d x},
\label{eq_chie}
\end{eqnarray}
where $x_{\rm in}$ and $x_{\rm out}$ represent the values of $x$ at inner and outer disk radius. In the case of $x_{\rm in}=0$ and $x_{\rm out}=\infty$, with the integral formula (\citealt{Stegun1972})
\begin{eqnarray}
\int_{0}^{\infty} \frac{x^{n}}{e^{x}-1 }dx=\zeta{(n+1)}\Gamma{(n+1)},
\end{eqnarray}
the emissivity-weighted factor can be analytically expressed as 
\begin{equation}
\chi_{\rm e} = \frac{\zeta{(3\beta)}\Gamma{(3\beta)}}{\zeta{(2\beta)}\Gamma{(2\beta)}},
\end{equation}
where $\zeta(x)$ and $\Gamma(x)$ are the zeta function and gamma function respectively \citep{Stegun1972}. As can be seen, the black hole mass and accretion rate do not appear in the far right-hand side of Equation~(\ref{eq_chie}). Therefore, $\chi_{\rm e}$ is independent on black hole mass and accretion rate.

Similarly, we can recast the expression for the responsivity-weighted factor $\chi_{\rm r}$ as 
\begin{eqnarray}
\chi_{\rm r} = \frac{1}{R_{\lambda}} \frac{\int_{R_{\rm in}}^{R_{\rm out}}\frac{\partial B_\lambda(T)}{\partial T}\frac{\partial T}{\partial f_c}R^2 d R}{\int_{R_{\rm in}}^{R_{\rm out}}\frac{\partial B_\lambda(T)}{\partial T}\frac{\partial T}{\partial f_c}R d R} 
= \frac{\int_{x_{\rm in}}^{x_{\rm out}} \frac{e^x}{(e^{x}-1)^2 }x^{3\beta}dx} {\int_{x_{\rm in}}^{x_{\rm out}}\frac{e^x}{(e^{x}-1)^2}x^{2\beta} d x}.
\end{eqnarray}
Again, with the aid of the zeta and gamma functions, we have 
\begin{eqnarray}
 \chi_{\rm r} = \frac{\zeta{(3\beta)}\Gamma{(3\beta+1)}}{\zeta{(2\beta)}\Gamma{(2\beta+1)}},
\end{eqnarray}
where we used the integral formula
\begin{eqnarray}
\int_{0}^{\infty} \frac{ e^{x}x^{n}}{(e^{x}-1)^2 }dx=\zeta{(n)}\Gamma{(n+1)}.
\end{eqnarray}

\begin{figure*}[t!]
    \centering 
    \includegraphics[width=0.45\textwidth]{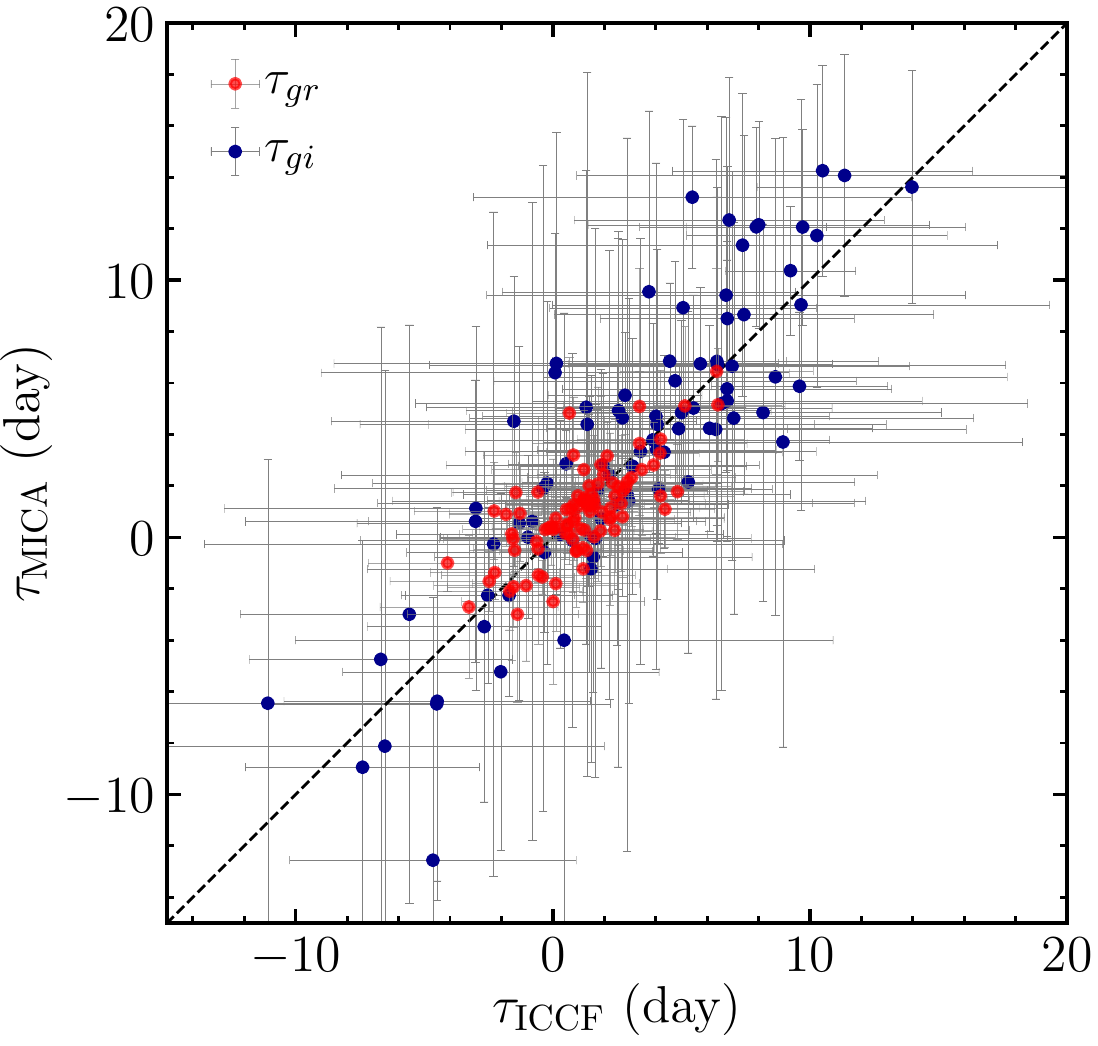}
    \includegraphics[width=0.45\textwidth]{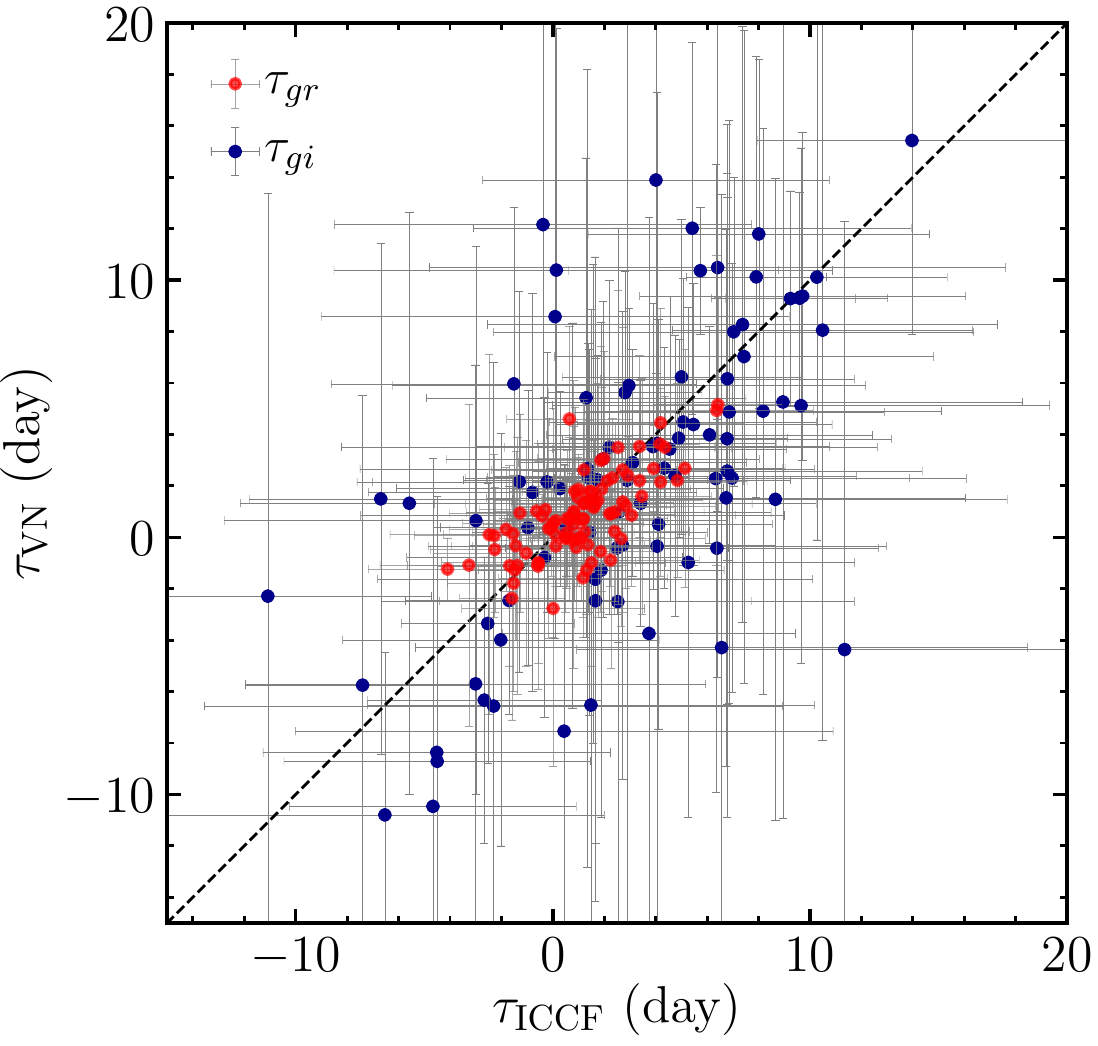}
    \caption{{ Comparisons of inter-band time delays measured by the ICCF method with (left) the MICA method and (right) von Neumann Esitmator.} The orange and blue points represent $\tau_{gr}$ and $\tau_{gi}$ (in the observed frame), respectively.}
    \label{fig_ccf_mica}
\end{figure*}

\section{Time Delay Measurements from the MICA and von Neumann Estimator } 
\label{sec_mica}
Figure~\ref{fig_ccf_mica} compares the measured inter-band time delays  from the ICCF method with those from the MICA method and the von Neumann estimator. We can find general consistency among the three approaches. The uncertainties of $\tau_{gr}$ obtained by MICA are a factor of 0.74 of those obtained by the ICCF method, whereas the uncertainties of $\tau_{gi}$ obtained by the two methods are comparable. The uncertainties from the von Neumann estimator are overall 1.32 times larger than those from the ICCF method.

\begin{figure}[t!]
\centering 
\includegraphics[width=0.7\textwidth]{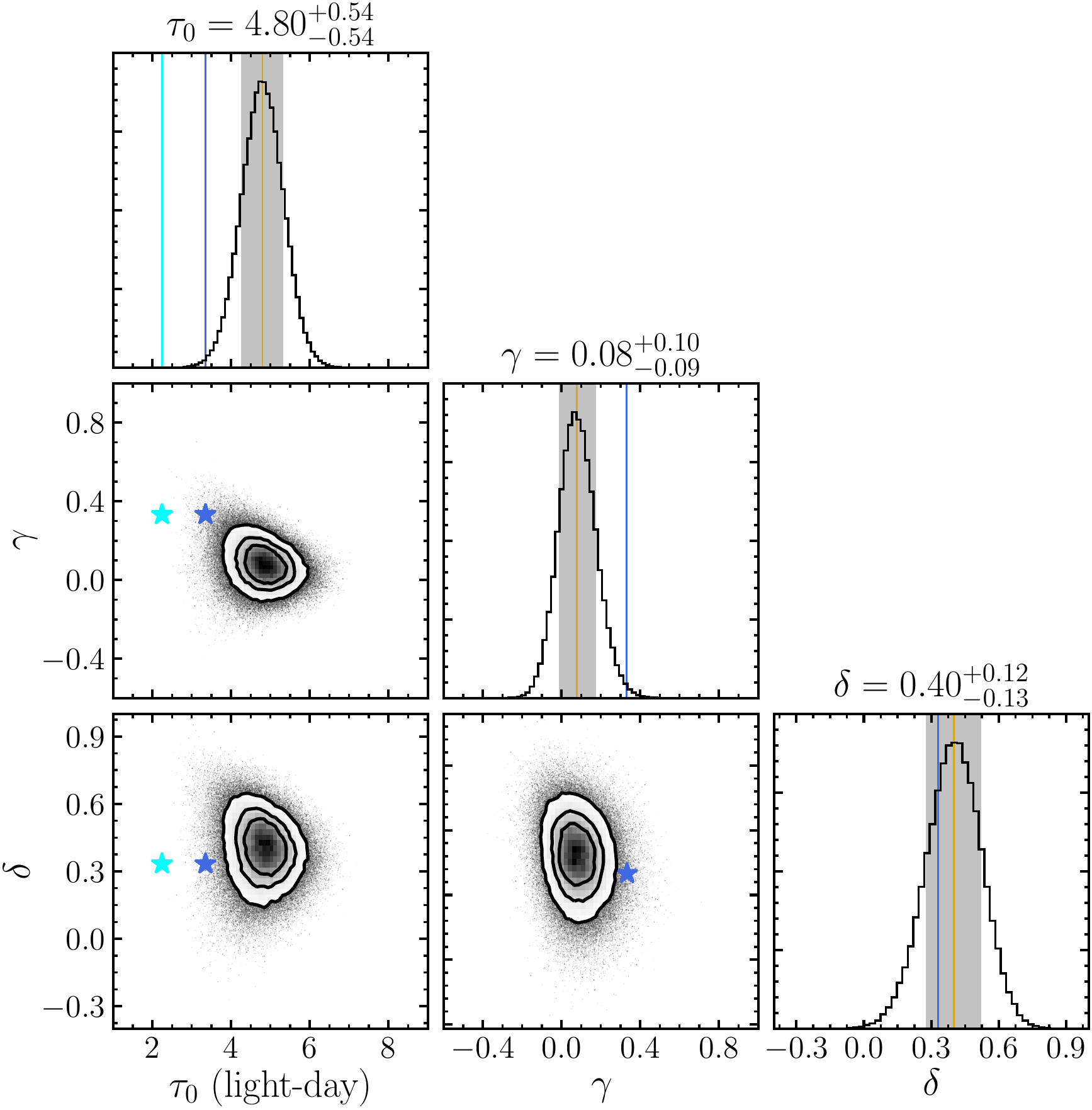}
\caption{Same as Figure~\ref{fig_caseIII}, but using the MICA-based inter-band time delays.}
\label{fig_caseI_mica}
\end{figure}

We also run the fitting analysis using time delays from MICA and von Neumann estimator.  Table \ref{tab_method} compares 
the obtained results from ICCF, MICA, and von Neumann estimator, which, again, are consistent within uncertainties among the three approaches.
In Figure~\ref{fig_caseI_mica}, we plot the fitting results of Case~III using the MICA-based time delays. As expected, the best estimates of parameter $\tau_0$, $\delta$, and $\gamma$ are consistent with the ICCF results but have relatively smaller uncertainties.

 \begin{deluxetable*}{lccccc}
\renewcommand\arraystretch{1.2}
\tablecolumns{5}
\tabletypesize{\footnotesize}
\tabcaption{ The best estimated disk parameters using ICCF, MICA, and von Neumann estimator. \label{tab_method}}
\tablehead{
\colhead{} &
\colhead{~~~~~~~Parameter~~~~~~~} &
\colhead{~~~~~~~ICCF~~~~~~~} &
\colhead{~~~~~~~~MICA~~~~~~~~~} &
\colhead{~~~~~~~~von Neumann Estimator~~~~~~~~} &
    \colhead{~~~  ~~~~~Theoretical Value~~~~~~~~}
}
\startdata    
Case I    & $\tau_0$ (light-day)  & $4.56_{-0.69}^{+0.69}$  & $4.23_{-0.43}^{+0.43}$ & $4.31_{-0.79}^{+0.79}$  & \ 2.24 (3.36)\\\hline
Case II     & $\tau_0$ (light-day)   &  $5.29_{-1.69}^{+4.05}$     & $5.12_{-1.15}^{+2.05}$  & $5.72_{-2.26}^{+5.97}$ & 2.24 (3.36) \\
 & $\beta$  &  $1.08_{-0.57}^{+0.61}$    & $1.02_{-0.36}^{+0.38}$ & $0.93_{-0.54}^{+0.69}$ & 4/3 \\\hline
Case III     & $\tau_0$ (light-day)   &  $4.68_{-0.91}^{+0.88}$    & $4.80_{-0.54}^{+0.54}$  & $4.55_{-1.04}^{+1.00}$ & 2.24 (3.36) \\
& $\gamma$ &  $0.06_{-0.15}^{+0.15}$    & $0.08_{-0.09}^{+0.10}$ & $0.05_{-0.17}^{+0.20}$ & 1/3 \\
& $\delta$ &  $0.44_{-0.22}^{+0.19}$   & $0.40_{-0.13}^{+0.12}$ & $0.35_{-0.25}^{+0.24}$&  1/3\\
\enddata
\tablecomments{Two theorectical values of $\tau_0$ represent the disk sizes using emissivity- and responsivity-weighted factor, respectively.}
\end{deluxetable*}

\bibliographystyle{aasjournal}
\bibliography{ref}
\end{document}